\documentclass[smallextended]{svjour3}
\smartqed 
\usepackage{amssymb,amsmath,graphicx}
\usepackage{amsfonts}
\usepackage{caption, subfigure}
\usepackage{epsfig,latexsym,graphicx}
\begin{document}


\title{{\bf Divergence based Robust Estimation of the Tail Index through An Exponential Regression Model}}

\author{ Abhik Ghosh       
}


\institute{A. Ghosh \at
              Interdisciplinary Statistical Research Unit\\
              Indian Statistical Institute, Kolkata, India.\\
              \email{abhianik@gmail.com}           
}

\date{Received: date / Accepted: date}

\maketitle

\begin{abstract}
The extreme value theory is very popular in applied sciences including finance, economics, 
hydrology and many other disciplines. In univariate extreme value theory, we model the data by a suitable distribution 
from the general max-domain of attraction (MAD) characterized by its tail index; 
there are three broad classes of tails -- the Pareto type, the Weibull type and the Gumbel type.
The simplest and most common estimator of the tail index is the Hill estimator 
that works only for Pareto type tails and has a high bias; 
it is also highly non-robust in presence of outliers with respect to the assumed model. 
There have been some recent attempts to  produce asymptotically unbiased or robust alternative to the Hill estimator;
however all the robust alternatives work for any one type of tail.  
This paper proposes a new general estimator of the tail index that is both robust and has smaller bias 
under all the three tail types compared to the existing robust estimators. This essentially produces 
a robust generalization of the estimator proposed by Matthys and Beirlant (2003) under the same 
model approximation through a suitable exponential regression framework using the density power divergence. 
The robustness properties of the estimator are derived in the paper 
along with an extensive simulation study.
A method for bias correction is also proposed with application to some real data examples.
\keywords{Extreme Value Theory \and Robust Methods \and Exponential Regression Model \and Density power divergence}
\end{abstract}

\section{Introduction}

The recent exploration in scientific technology and modern instruments have expanded 
the scope of research in all fields of life. This has increased the need for 
suitable analytical techniques to ensure 
the quality of overloaded datasets in the laboratory.
In a variety of applied disciplines including economics, finance, hydrology etc., 
any decision obtained from statistical modeling based on those datasets leads to 
a new innovation in the respective fields at  a huge cost and 
hence the investment has to be insured beforehand carefully to the extent possible 
against its potential adverse effects. 
Therefore, risk management has become a very important area of research in the recent era 
and arguably the most difficult area of it is to model the very rare but dangerous events 
that produce huge risk (loss) known in practice as the ``worst-case risk".  
These events arise in analyzing unusual big claims in insurances, 
studying equity risk, predicting rare natural disasters etc.
These problems cannot be solved using the regular normal models; 
the statistical framework that helps to analyze such situations is the extreme value theory.
The extreme value models generally have a thicker tail compared to the normal models
and the probabilities of rare events are modeled by the non-zero tail probabilities
of such heavy-tailed distributions. 
For any univariate distribution, such a tail behavior is characterized by 
its tail index that measures, in a layman's term, the thickness of the tail.

In terms of the statistical terminology, let $X_1, \ldots, X_n, \ldots$ 
denote independent and identically distributed data on some natural process 
like daily stock returns 
and we model these observations by a distribution function $F$ having density $f$.
Then, the probability of any extreme event can be found by estimating the quantity
$\bar{F}(x) = 1 - F(x)= P(X_i > x)$ for some large threshold $x$. 
In order to infer about the extreme events beyond the sample range,
one assumes that the distribution of sample maximum $X_{(n)} = \max\{X_1, \ldots, X_n\}$
(properly standardized) converges to a non-degenerate distribution indexed by 
a parameter $\gamma$ (say) known as the tail index of the distribution $F$.
More precisely, following Gnedenko (1943) one assumes the existence of two sequences of constants 
$\{a_n\}>0$ and $\{b_n\} \subset \mathbb{R}$ satisfying
\begin{eqnarray}
\label{EQ:EVT_def1}
\lim\limits_{n \rightarrow \infty}~ P\left(\frac{X_{(n)}-b_n}{a_n} \leq x\right) = H_\gamma(x),
\end{eqnarray}
for all continuity points $x$ of the extreme value distribution $H_\gamma(x)$.
Then the distribution $F$ of the original sample is said to belong to 
the maximum domain of attraction (MAD) of  $H_\gamma$ and 
can be classified into three board classes:
\begin{enumerate}
\item Fr\'{e}chet class of distributions with $\gamma>0$ : Pareto, Burr, Student's t, log-gamma etc.,
all having slowly decaying tails;
\item Gumbel class of distributions with $\gamma=0$ : Exponential, Weibull, normal, Gamma, lognormal etc.,
all having tails decaying exponentially fast;
\item Weibull class of distributions with $\gamma<0$ : Uniform, reversed Burr, Beta, reversed Pareto etc.,
all having finite right tails.
\end{enumerate}

Estimation of the tail index $\gamma$ is 
the main problem in extreme value theory and, 
as one can expect, there is a large literature which deals with the same.
In this regard, the simplest classical estimator of the tail weight $\gamma$ is 
Hill's (1975) estimator defined by 
\begin{equation}
\widehat{\gamma}_H = \frac{1}{k} ~\sum_{i=1}^k\log(X_{(n-i+1)}) - \log(X_{(n-k)}),
\label{EQ:Hill_est}
\end{equation}
where $X_{(i)}$ denotes the $i^{\rm th}$ order statistics in $\{X_1, \ldots, X_n\}$
and $k$ is the number of extreme observations to be used.
Although Hill's estimator is very popular in extreme value theory, 
it only works under the Pareto type tails with $\gamma>0$. 
Smith (1987) derived a maximum likelihood estimator of tail index 
using the generalized Pareto distribution for excess over a high threshold (POT) 
that has a non-degenerate asymptotic distribution for $\gamma> - 1/2$.
On the other end, Hosking and Wallis (1987) derived an estimator of  the tail index 
that gives good results for $\gamma<1$. The estimation of all the three types of tail indices 
was proposed by Pickands (1975) although that was latter found to be 
unstable with respect to the choice of the sample proportion used $(k/n)$.
The moment type estimator of the general $\gamma\in \mathbb{R}$, 
proposed by Dekkers et al.~(1989), has become popular due to its simple interpretation,
although it has quite high asymptotic variance for negative $\gamma$.
Beirlant et al.~(1999) developed 
a maximum likelihood estimator of the tail index based on 
an exponential regression model approximation for $\gamma>0$, 
which has been extended by Matthys and Beirlant (2003) for all the three types of tails.
The latter method has an asymptotic variance 
which is smaller compared to the variance of the moment  type estimator at $\gamma<0$
and almost equal to the variance of the POT estimator at $\gamma>0$.

However, the existing literature referred to  above does not take into account 
the possible outlying observations present in the sample and most of these estimators, 
if not all, are highly sensitive to such outliers. 
However, in real practice, there could be a significant portion 
of outliers in the datasets with respect to the assumed model 
either due to ignorance of some external factors 
or erroneous input at some level of data collection.  
The inference about the tail events using such observations generates 
incorrect insights producing a big loss as mentioned earlier.
However, in most cases, it is not easy to separate out those outliers from a large dataset beforehand.
Thus, automatic outlier control with some robust tool
is very crucial to manage the quality of data and the overall inference.
This part was completely ignored previously due to the prior conception that 
the two theory of extreme value statistics and robust statistics are contradictory 
as the first one models the large observations in the sample and the second ignores them. 
That there could be two types of large observations in a sample
and need to be handled separately to get more accurate results 
have been noticed more recently and some attempts have been made to produce robust estimators 
of the tail index. These include Vandewalle et al.~(2004, 2007) and  Kim and Lee (2008);
but these estimators are proposed and studied only for the Pareto type tails with $\gamma>0$.
Recently Goegebeur et al.~(2014) have derived a robust estimator for a class of Weibull-type distributions, 
a subset of the Gumbel max-domain of attraction having $\gamma=0$, 
using the conditional approach with some covariates.
However, in practice prior knowledge about the type of tail is often not available
since it is difficult to figure out from the data alone before estimating the tail index $\gamma$.
Therefore, a robust estimator simultaneously considering all the three types of tails  
[like the non-robust estimator of Matthys and Beirlant (2003)]
would be really helpful to a wide range of practitioners in several applied fields  
like risk management, finance, hydrology  and many others.

The present paper aims to propose one such estimator extending the 
concept of Matthys and Beirlant (2003); we will use the robust minimum 
density power divergence estimation technique in place of the non-robust 
maximum likelihood method. The minimum density power divergence estimator, 
proposed by Basu et al.~(1998), has become a very popular robust alternative to 
the maximum likelihood estimator in recent times and has the advantage of 
high robustness with only a small loss in asymptotic efficiency; 
further the estimation process is no more complicated than 
the maximum likelihood estimation. 
The density power divergence down-weights the outliers by a non-zero power of the model density
and we will exploit this fact to derive a robust estimator of the tail index
under the exponential regression model approximation to 
the log-ratio of ordered excess over a large threshold.

The rest of the paper is organized as follows: We start with a brief description
of the maximum likelihood estimator of $\gamma$ under the exponential regression model (ERM) 
from Matthys and Beirlant (2003) in Section \ref{SEC:MLE_ERM}
to understand the model conditions and notations more clearly.
In Section \ref{SEC:MDPDE_ERM} we will present the robust estimator of 
the tail index by minimizing the density power divergence 
between data and the approximated exponential regression model; 
we will also prove their robustness for all the three types of tails.
Then the performance of the proposed estimator will be illustrated 
through an extensive simulation study in Section \ref{SE:Simulation}.
Section \ref{SEC:choice_parameter} will present some discussion on 
the source of bias and the choice of tuning parameters. 
Section \ref{SEC:Bias_corr} will present an approach for correcting that bias with applications to some 
interesting real data examples.
Finally we will conclude this paper with some remarks in Section \ref{SEC:cnclusion}.

\section{The Exponential Regression Model for Tail Index Estimation and the Non-robust Maximum Likelihood}
\label{SEC:MLE_ERM}

Let us consider a random sample $X_1, \ldots, X_n$ from a distribution $F$ 
that belongs to the maximum domain of attraction of the extreme value distribution $H_\gamma$.
Therefore $F$ satisfies Condition (\ref{EQ:EVT_def1}) that was reformulated 
by an equivalent condition in de Haan (1970); the latter assumes the existence of 
a measurable positive function $a_Q(\cdot)$ such that for all $\lambda>0$
\begin{eqnarray}
\lim\limits_{t\rightarrow\infty} \frac{Q(\lambda t) - Q(t)}{a_Q(t)} = 
\left\{\begin{array}{l c l}
\frac{\lambda^\gamma-1}{\gamma} & \mbox{ for } & \gamma \ne 0,\\
\log \lambda & \mbox{ for } & \gamma = 0,
\end{array}\right.
\label{EQ:EVT_def2}
\end{eqnarray}
where $Q$ is the tail quantile function defined by 
$Q(t) = \inf\left\{x : F(x) \geq 1 - \frac{1}{t}\right\}.$
This condition helps us to derive an useful nonparametric approximation to the 
ordered spacing of the observed sample as shown in Matthys and Beirlant (2003).
Let $X_{(1)} \leq X_{(2)} \leq \cdots \leq X_{(n)}$ denote the ordered sample;
$U_{(1)} \leq U_{(2)} \leq \cdots \leq U_{(n)}$ denote the order statistics from $n$ i.i.d.~uniform(0,1) observations,
$V_{(1)} \leq V_{(2)} \leq \cdots \leq V_{(k)}$ and $E_{(1)} \leq E_{(2)} \leq \cdots \leq E_{(k)}$ 
denote the same from  $k$ i.i.d.~uniform(0,1) and exponential(1) random variables respectively.
Throughout this paper, we will write $W \mathop{=}^d Z$ to mean that $W$ and $Z$ have the same distribution 
and $W \mathop{\sim}^d Z$ to mean that they have the same asymptotic distribution.

Now for a fixed $k<n$ and any $j=1, \ldots, k$, we get
\begin{eqnarray}
X_{(n-j+1)} - X_{(n-k)} &\mathop{=}^d& Q(U_{(j)}^{-1}) - Q(U_{(k+1)}^{-1}) \nonumber\\
&\mathop{=}^d& Q(U_{(k+1)}^{-1}V_{(j)}^{-1}) - Q(U_{(k+1)}^{-1}) \nonumber\\
&\mathop{\sim}^d& a_Q(U_{(k+1)}^{-1}) \frac{V_{(j)}^{-\gamma} - 1}{\gamma},
~~~~~[\mbox{by Condition (\ref{EQ:EVT_def2})}]. \nonumber
\end{eqnarray}
Then, we have 
\begin{eqnarray}
\log\left(\frac{X_{(n-j+1)} - X_{(n-k)}}{X_{(n-j)} - X_{(n-k)}}\right)
&\mathop{\sim}^d&  \log\left(\frac{V_{(j)}^{-\gamma} - 1}{V_{(j+1)}^{-\gamma} - 1}\right) \nonumber\\
&\mathop{=}^d&  \log\left(e^{\gamma E_{(k-j+1)}} - 1\right) - \log\left(e^{\gamma E_{(k-j)}} - 1\right) \nonumber\\
&\mathop{=}^d&  \left(E_{(k-j+1)} - E_{(k-j)}\right) \frac{\gamma e^{\gamma E^*}}{e^{\gamma E^*}-1} \nonumber\\
&&[\mbox{by the Mean Value Theorem (MVT);}\nonumber\\
&& ~\mbox{ $E^*$ in between $E_{(k-j)}$ and $E_{(k-j+1)}$}] \nonumber \\
&\mathop{=}^d&  \frac{E_{k-j+1}}{j} \cdot \frac{\gamma }{1 - \left(e^{- E^*}\right)^\gamma}, \nonumber
\end{eqnarray}
where $E_1, \ldots, E_k$ are $k$ i.i.d.~observations from an exponential distribution with mean $1$.
This set of equations follows by the Renyi representation; see Matthys and Beirlant (2003) for more details.
Now the quantity $\left(e^{- E^*}\right)$ lies in between $V_{(k-j)}$ and $V_{(k-j+1)}$ and 
so it can be estimated by $\frac{j}{k+1}$. 
Hence, we get an exponential regression model approximation for the scaled log-ratios 
of ordered spacing given by 
\begin{eqnarray}
j\log\left(\frac{X_{(n-j+1)} - X_{(n-k)}}{X_{(n-j)} - X_{(n-k)}}\right)
&\mathop{\sim}^d&  \frac{\gamma }{1 - \left(\frac{j}{k+1}\right)^\gamma} E_{k-j+1}, ~~ j = 1, \ldots, k-1.
\label{EQ:ERM_def1} 
\end{eqnarray}
Let us denote the left hand side of Equation (\ref{EQ:ERM_def1}) by $Y_j$ 
for $j = 1, \ldots, k-1$. Then, asymptotically the distribution of $Y_j$ is 
exponential with mean $\theta_j =  \frac{\gamma }{1 - \left(\frac{j}{k+1}\right)^\gamma}$
which can be used to estimate the tail index $\gamma$.
One important advantage of the above construction is that the values of $Y_j$s remain
invariant under location and scale transformation of the data and so will be the 
corresponding estimator of $\gamma$ obtained using $Y_j$s. 
So, the estimator of tail index will be independent of the measurement unit of the data.

Matthys and Beirlant (2003) proposed to estimate the tail index $\gamma$ by maximizing 
the log-likelihood corresponding to the above exponential regression model given by 
$$
l(\gamma) = \sum_{j=1}^{k-1}~\left[\log\left(\frac{1 - \left(\frac{j}{k+1}\right)^\gamma}{\gamma}\right) 
- Y_j \left(\frac{1 - \left(\frac{j}{k+1}\right)^\gamma}{\gamma}\right) \right].
$$
Differentiating the above with respect to $\gamma$, the maximum likelihood estimator of $\gamma$,
denoted by $\hat{\gamma}_{MLE}$, can be obtained as a solution of the estimating equation
\begin{eqnarray}
\sum_{j=1}^{k-1}~\widetilde{J}\left(\frac{j}{k+1}\right) 
\left[Y_j  - \frac{\gamma}{1 - \left(\frac{j}{k+1}\right)^\gamma}\right]=0,
\label{EQ:Est_eqn_MLE}
\end{eqnarray} 
where $\widetilde{J}(u) = (u^\gamma -1 - \gamma u^\gamma \log u)/\gamma^2$.
The asymptotic distribution and consistency of $\hat{\gamma}_{MLE}$ are derived in 
Matthys and Beirlant (2003)  under suitable assumptions. 
Further it has been observed that, in terms of asymptotic variance,
$\hat{\gamma}_{MLE}$ performs comparably  to the POT estimator at $\gamma>0$ 
and significantly better compared to the moment type estimators at $\gamma<0$; 
at $\gamma=1$ all the three estimators have equal asymptotic variance $1$.
However, in spite of having asymptotically optimal properties, 
the crucial problem of any maximum likelihood estimator is the lack of robustness 
with respect to the outlying observation in the sample.
So $\hat{\gamma}_{MLE}$ is also highly non-robust with respect to outliers and 
in this paper we will present a robust generalization of this estimator using 
the density power divergence.

\section{Robust Estimation of the Tail Index through the ERM by minimizing the Density Power Divergence}
\label{SEC:MDPDE_ERM}

The density power divergence (DPD), proposed by Basu et al.~(1998), 
has become very popular now-a-days in the context of robust inference.
It uses the philosophy of weighted likelihood estimating equation, 
where the outlying observations having low model probabilities are 
down-weighted by a non-zero power $\alpha$ of the model density.
Thus, the density power divergence is defined in terms of the tuning parameter $\alpha$ 
as follows:
\begin{equation}
d_\alpha(g,f_\theta) = \int  f_\theta^{1+\alpha} - \frac{1+\alpha}{\alpha} \int f_\theta^\alpha g 
+ \frac{1}{\alpha} \int g^{1+\alpha}, ~~~~~\mbox{ if } ~~ \alpha > 0. \nonumber 
\end{equation}
For $\alpha = 0$, the corresponding divergence can be defined as the continuous limit of
the above divergence as $\alpha \downarrow 0$, which is nothing but the 
Kulback-Leibler divergence: 
\begin{equation} 
 d_0(g,f_\theta) = \lim_{\alpha \rightarrow 0} d_\alpha(g,f_\theta) = \int g \log(g/f_\theta). \nonumber
\end{equation}
For independent and identically distributed sample $X_1, \ldots, X_n$ from a population 
to be modeled by a parametric family $\{f_\theta:\theta \in \Theta\}$, 
the minimum density power divergence estimator (MDPDE) of the parameter of interest $\theta$ 
has to be obtained by minimizing the divergence between the data and model density,
or equivalently by minimizing the quantity 
$$
H_n(\theta) = \int~f_\theta^{1+\alpha} - \frac{1+\alpha}{\alpha} \frac{1}{n} \sum_{i=1}^n f_\theta^\alpha(X_i)
$$
with respect to $\theta \in \Theta$. Under suitable assumptions, 
the MDPDE of $\theta$ can be seen to be consistent and asymptotically normal.
Further, Basu et al.~(1998) have shown that the tuning parameter $\alpha$ controls 
the  trade-off between asymptotic efficiency and robustness; 
at $\alpha=0$ we have the most efficient but highly non-robust maximum likelihood estimator (MLE)
and at $\alpha=1$ it coincides with the $L_2$-divergence generating a highly robust but relatively inefficient estimator. 
They have also argued that the consideration of MDPDE with $\alpha>1$ is unnecessary;
in fact MDPDEs with a small positive $\alpha$ gives quite satisfactory robust results 
with a very little loss in efficiency.

Here we want to obtain the minimum DPD estimator of the tail index
based on an observed sample $X_1, \ldots, X_n$ from a population having distribution function $F$. 
However, we have not assumed any parametric model for the corresponding population
and transform the data from $X_1, \ldots, X_n$ to $Y_1, \ldots, Y_{k-1}$ 
as defined in the previous section with $k$ being the number of extreme observations to be used.
Then, as we have seen, with only the assumptions of $F\in MAD(H_\gamma)$, 
we can approximate the distribution of the transformed observations $Y_j$s by 
a suitable exponential regression model. Note that the transformed sample observations $Y_j$
are no-longer identically distributed, although they are still independent. 
Thus, we cannot directly apply the original formulation of the MDPDE as described above; 
we need a suitable generalization for non-homogeneous data.
Ghosh and Basu (2013) provide one such generalization by minimizing the average 
DPD measures computed separately for all the sample points.
In this paper, we will follow the approach of Ghosh and Basu (2013) to 
produce a robust estimator of the tail index.

\subsection{Estimating Equation}

Consider the set-up of Section \ref{SEC:MLE_ERM} with a random sample 
$X_1, \ldots, X_n$ from the distribution $F\in MAD(H_\gamma)$. 
Define $Y_j = j\log\left(\frac{X_{(n-j+1)} - X_{(n-k)}}{X_{(n-j)} - X_{(n-k)}}\right)$;
let its true distribution and density functions be $G_j$ and $g_j$ respectively (obtained from $F$).
As argued in the previous section, we will model this by 
an exponential regression model (\ref{EQ:ERM_def1}) so that $Y_j$s independently follows $f_{\theta_j}$, 
where $f_\theta$ is the exponential density with mean $\theta$. 
Note that $\theta_j$ is a non-linear function of the parameter of interest $\gamma$.
Then following Ghosh and Basu (2013), the MDPDE
of $\gamma$ has to be obtained by minimizing the average discrepancy
\begin{eqnarray}
H_k(\gamma) = \frac{1}{k-1} \sum_{j=1}^{k-1} ~ \left[\int~f_{\theta_j}^{1+\alpha} 
- \frac{1+\alpha}{\alpha} \int f_{\theta_j}^\alpha\hat{g}_j\right] ,
\label{EQ:MDPDE_est_eqn0}
\end{eqnarray}
where $\hat{g}_j$ is some non-parametric estimator of $g_j$ based on the observed sample.
Note that, here  the size of the transformed sample is $k-1$; 
we will assume that as the original sample size $n\rightarrow \infty$, 
$k$ also tends to infinity.
Further, for each $j$ we have only one observation $Y_j$ from $g_j$
and so  the best possible non-parametric estimator of $g_j$ 
is given by the degenerate distribution at $Y_j$. Then, using the form of exponential density, 
the above objective function can be seen to have the form:
\begin{eqnarray}
H_k(\gamma) = \frac{1}{k-1} \sum_{j=1}^{k-1} ~ \left[\frac{1}{(1+\alpha)\theta_j^\alpha} - 
\frac{(1+\alpha)}{\alpha\theta_j^\alpha} e^{-\frac{\alpha Y_j}{\theta_j}} \right],
\label{EQ:MDPDE_obj_func}
\end{eqnarray}

Equivalently, we can obtain the MDPDE of the tail index $\gamma$ 
by solving the estimating equation $\nabla_\gamma H_k(\gamma) = 0$, 
where $\nabla_\gamma$ represents the first order partial derivative with respect to $\gamma$.
A routine differentiation of (\ref{EQ:MDPDE_obj_func}) yields 
the following simplified form of the estimating equation:
\begin{eqnarray}
\sum_{j=1}^{k-1}  \widetilde{J}_\alpha\left(\frac{j}{k+1}\right)  \left[\frac{\alpha\theta_j}{(1+\alpha)^2} + 
\left(Y_j - \theta_j\right) e^{-\frac{\alpha Y_j}{\theta_j}} \right] =0, 
\label{EQ:MDPDE_est_eqn}
\end{eqnarray} 
where $\widetilde{J}_\alpha(u) = (u^\gamma -1 - \gamma u^\gamma \log u)(1-u^\gamma)^\alpha\gamma^{-\alpha-2}$. 
Whenever the above estimating equation has more than one root, 
we choose the one that minimizes the objective function $H_k(\gamma)$.
We will denote the corresponding minimum DPD estimator 
of the tail index $\gamma$ by $\hat{\gamma}_{ER,k}^{(\alpha)}$, 
where $k$ and $\alpha$ are the tuning parameters used.
Interestingly, the above MDPDE estimating equation (\ref{EQ:MDPDE_est_eqn}) 
coincides with the maximum likelihood estimating equation (\ref{EQ:Est_eqn_MLE}) at $\alpha=0$ and 
hence $\hat{\gamma}_{ER,k}^{(0)}$ is nothing but the estimator proposed in Matthys and Beirlant (2003),
to be denoted by ``MB estimator" throughout the rest of this paper.
Since the case $\alpha=0$ provides no outlier down-weighting,
the corresponding estimator is clearly non-robust; 
the MDPDEs with $\alpha>0$
provide its robust generalization. 
In the next subsection, we will rigorously examine their robustness through the 
classical influence function analysis.

\subsection{Robustness: Influence Function Analysis}

The most common and classical tool for measuring robustness is 
Hampel's (1968, 1974) influence function.
It indeed gives us the first order approximation to the asymptotic 
bias of the estimator under infinitesimal contamination at an outlying point in the sample space.
So whenever the influence function of an estimator is bounded its bias cannot increase
indefinitely even if there is a strong contamination in a point 
far away from the central cloud of the model distribution.
The supremum of the influence function over all possible outlier points 
yields a measure of the extent of robustness of the estimator with small values being preferred.
We will now derive the influence function of the proposed MDPDE
of the tail index under the exponential regression model approximation.

In order to obtain the influence function, we need to re-define the estimator 
$\hat{\gamma}_{ER,k}^{(\alpha)}$ in terms of a statistical functional. 
For simplicity, we will work with the transformed variables $Y_j$, $j=1, \ldots, k-1$
and let $\underline{\mathbf{G}}=(G_1, \cdots, G_{k-1})$.
Following Equation (\ref{EQ:MDPDE_est_eqn0}), it can be seen that 
$\hat{\gamma}_{ER,k}^{(\alpha)} = T_\alpha(\underline{\mathbf{\hat{G}}})$ 
where $\hat{G}_j$ denotes the distribution function of $\hat{g}_j$ and 
the functional $T_\alpha(\underline{\mathbf{{G}}})$ is defined as the minimizer of 
\begin{eqnarray}
&& \frac{1}{k-1} \sum_{j=1}^{k-1} ~ \left[\int~f_{\theta_j}^{1+\alpha} 
- \frac{1+\alpha}{\alpha} \int f_{\theta_j}^\alpha{g}_j\right] \nonumber\\
&=& \frac{1}{k-1} \sum_{j=1}^{k-1} ~ \left[\frac{1}{(1+\alpha)\theta_j^\alpha} - 
\frac{(1+\alpha)}{\alpha\theta_j^\alpha} \int e^{-\frac{\alpha y}{\theta_j}} g_j(y)dy \right],\nonumber
\end{eqnarray}
with respect to $\gamma$. Note that the statistical functional corresponding to 
the estimator $\hat{\gamma}_{ER,k}^{(\alpha)}$ depends on the parameter $k$ 
as in the case of Ghosh and Basu (2013) for non-homogeneous observations.  
Therefore, the corresponding influence function will also  depend on $k$, 
the number of extreme sample observations to be used in estimation; so we will refer to it 
as the fixed-sample influence function. Let $\gamma^g = T_\alpha(\underline{\mathbf{{G}}})$ 
be the true best fitting parameter value. Then, by using the divergence property of the DPD, 
one can show that the statistical functional $T_\alpha$ is Fisher consistent.
Note that the functional $T_\alpha(\underline{\mathbf{{G}}})$ satisfies the 
estimating equation
\begin{eqnarray}
\sum_{j=1}^{k-1}  \widetilde{J}_\alpha\left(\frac{j}{k+1}\right) 
\left[\frac{\alpha\theta_j}{(1+\alpha)^2} + 
\int ~ \left(y  - \theta_j\right)
e^{-\frac{\alpha y}{\theta_j}}g_j(y) dy \right]= 0. ~~
\label{EQ:MDPDF_est_eqn}
\end{eqnarray}

Now let us consider the contamination over the true distributions. 
Note that any contamination in our original sample $X_i$s with true distribution $F$ 
induces some amount of contamination in the transformed variables $Y_j$s having true distribution $G_j$.
Since we study the asymptotic effect of infinitesimal contamination (with contamination proportion $\epsilon \downarrow 0$),
the robustness properties will be the same even if we work with contamination in $Y_j$s instead of the actual observations $X_i$s
(only the form of contamination will differ but both will vanish as $\epsilon \downarrow 0$).
So, for simplicity, in this paper we will consider contamination in $Y_j$s (which is induced by some contamination in $X_i$s).
However, since $Y_j$s are not identically distributed, 
we have to consider contamination in each $Y_j$ separately as in Ghosh and Basu (2013).
Depending on the amount of contamination in $X_i$s there can be contamination in any particular $Y_j$ or in all $Y_j$s.

Let us first consider the simplest case where there is contamination only in one particular $Y_j$,
say in $Y_{j_0}$ for some $j_0 \in \{1, \ldots, k-1\}$.
Then we consider the corresponding contaminated distribution 
$G_{j_0, \epsilon} = (1-\epsilon) G_{j_0} + \epsilon \wedge_{t_0}$, 
where $\epsilon$ is the contamination proportion and $\wedge_{t_0}$ denotes 
the degenerate distribution at the contamination point $t_0$.
Define $\gamma_{\epsilon,j_0} = T_\alpha(G_1, \cdots, G_{j_0, \epsilon}, \cdots, G_{k-1})$ 
which should satisfy the estimating equation (\ref{EQ:MDPDF_est_eqn}) with 
$G_{j_0}$ replaced by $G_{j_0, \epsilon}$. Differentiating the resulting equation 
with respect to $\epsilon$ at $\epsilon=0$, or using the results of Ghosh and Basu (2013), 
we get the fixed-sample influence function of $T_\alpha$ based on $k$ extremes 
at the true distribution $\underline{\mathbf{G}}$ as follows:
\begin{eqnarray}
&& IF_{k,j_0}(t_0; T_\alpha, \underline{\mathbf{G}}) = 
\left.\frac{\partial \gamma_{\epsilon,j_0}}{\partial\epsilon}\right|_{\epsilon=0} \nonumber\\
&=&  \frac{\Psi_n^{-1}}{k-1} \widetilde{J}_\alpha\left(\frac{j_0}{k+1}\right)
\left[(t_0 - \theta_{j_0})e^{-\frac{\alpha t_0}{\theta_{j_0}}} -
\int (y - \theta_{j_0})e^{-\frac{\alpha y}{\theta_{j_0}}}g_{j_0}(y)dy \right],
\label{EQ:IF_MDPDE_cont1}
\end{eqnarray}
where $\gamma=\gamma^g$ and $\Psi_n$ is defined as in Equations (3.3) and (3.5) of Ghosh and Basu (2013).
Note that, this influence function has been derived for any underlying true distribution of $Y_j$s and 
so holds true for any distribution of $X_i$s. 
Also, for any such distribution, if follows from the boundedness of the function $se^{-s}$ that 
the above influence function of the proposed estimator $T_\alpha$ is bounded over the contamination point $t_0$
for all $\alpha>0$ and any $k$; this implies the robustness of our proposal for $\alpha>0$.
However, at $\alpha=0$ the above influence function becomes linear in $t_0$, and hence unbounded, 
implying the non-robust nature of the corresponding MB estimator 
even under contamination in one transformed variable.

For the purpose of illustration, we will simplify this expression for a particular case where the exponential regression model 
approximation (\ref{EQ:ERM_def1}) holds well enough so that we can replace $G_j$ 
by corresponding exponential distribution $F_{\theta_j}$ with $\gamma^g = \gamma$ in the above.
Denote $\underline{\mathbf{F}} = (F_{\theta_1},\cdots, F_{\theta_{k-1}})$.
Then, the fixed-sample influence function of the MDPDE of $\gamma$ at the model becomes 
\begin{eqnarray}
&& IF_{k,j_0}(t_0; T_\alpha, \underline{\mathbf{F}}) \nonumber\\
&=& \frac{(1+\alpha)^3}{(1+\alpha^2)}  
\left[\sum_{j=1}^{k-1}\frac{1}{\theta_j^{\alpha-2}}\widetilde{J}\left(\frac{j}{k+1}\right)^2\right]^{-1}
\widetilde{J}_\alpha\left(\frac{j_0}{k+1}\right)
\left[(t_0 - \theta_{j_0})e^{-\frac{\alpha t_0}{\theta_{j_0}}} + \frac{\alpha\theta_{j_0}}{(1+\alpha)}\right].~~
\nonumber
\end{eqnarray}
Clearly the MDPDE of the tail index $\gamma$ with $\alpha>0$
will be robust with respect to outliers at any particular  $Y_j$s for any choice of $k$. 
However, at $\alpha=0$ the influence function of the corresponding MDPDE, 
which is the same as the MB estimator, is given by 
\begin{eqnarray}
IF_{k,j_0}(t_0; T_0, \underline{\mathbf{F}}) &=& 
\left[\sum_{j=1}^{k-1}\widetilde{J}\left(\frac{j}{k+1}\right)^2\theta_j^2\right]^{-1}
\widetilde{J}\left(\frac{j_0}{k+1}\right)\left(t_0 - \theta_{j_0}\right),
\label{EQ:IF0_MDPDE0_cont1}
\end{eqnarray} 
which is a straight line with respect to $t_0$ and hence is unbounded; 
this clearly demonstrates the non-robust nature of the MB estimator.

\begin{figure}[h!t]
	\centering
	\subfigure[$\gamma=1$, $j_0 = 20$]{
		\includegraphics[width=0.3\textwidth] {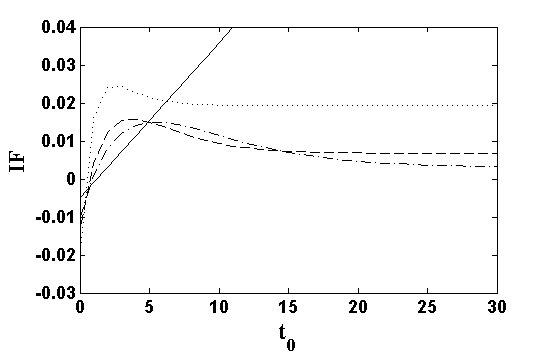}
		\label{FIG:IF_k100_g1p_J20}}
	~ 
	\subfigure[$\gamma=1$, $j_0 = 50$]{
		\includegraphics[width=0.3\textwidth] {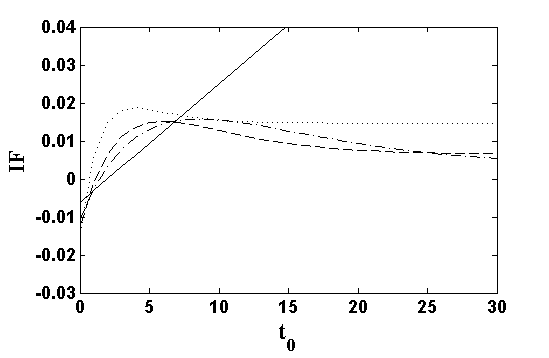}
		\label{FIG:IF_k100_g1p_J50}}
	~ 
	\subfigure[$\gamma=1$, $j_0 = 70$]{
		\includegraphics[width=0.3\textwidth] {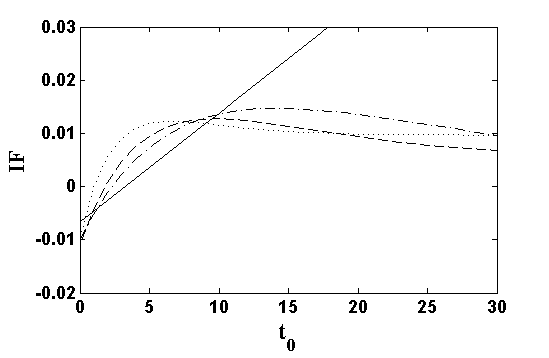}
		\label{FIG:IF_k100_g1p_J70}}
	\\ 
	\subfigure[$\gamma=0$, $j_0 = 20$]{
		\includegraphics[width=0.3\textwidth] {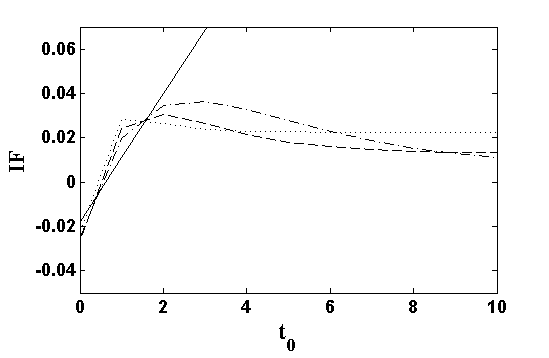}
		\label{FIG:IF_k100_g0_J20}}
	~ 
	\subfigure[$\gamma=0$, $j_0 = 50$]{
		\includegraphics[width=0.3\textwidth] {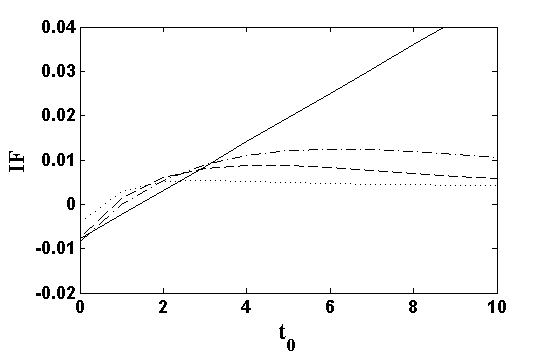}
		\label{FIG:IF_k100_g0_J50}}
	~ 
	\subfigure[$\gamma=0$, $j_0 = 70$]{
		\includegraphics[width=0.3\textwidth] {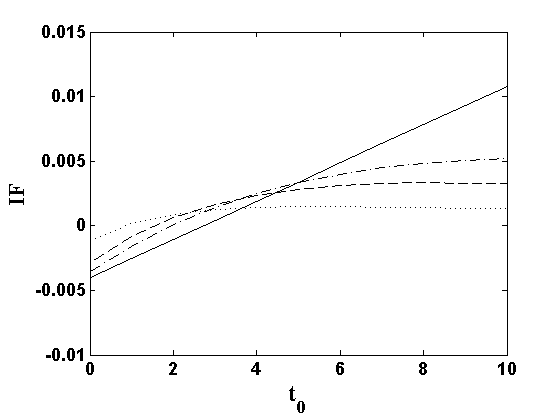}
		\label{FIG:IF_k100_g0_J70}}
	\\ 
	\subfigure[$\gamma=-1$, $j_0 = 20$]{
		\includegraphics[width=0.3\textwidth] {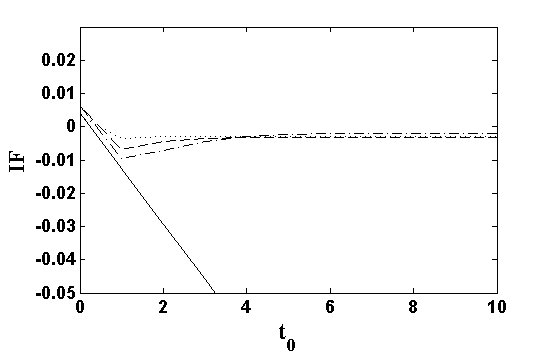}
		\label{FIG:IF_k100_g1n_J20}}
	~ 
	\subfigure[$\gamma=-1$, $j_0 = 50$]{
		\includegraphics[width=0.3\textwidth] {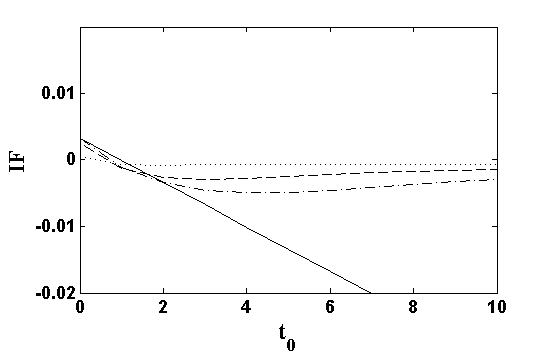}
		\label{FIG:IF_k100_g1n_J50}}
	~ 
	\subfigure[$\gamma=-1$, $j_0 = 70$]{
		\includegraphics[width=0.3\textwidth] {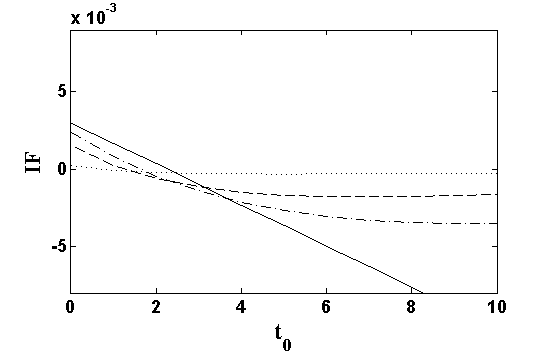}
		\label{FIG:IF_k100_g1n_J70}}
	\caption{Fixed-sample Influence Function of $T_\alpha$ over the contamination point for different types of tails with $k=100$ [Solid line: $\alpha=0$, Dotted line: $\alpha=0.3$, Dashed-dotted  line: $\alpha=0.5$, Dashed line: $\alpha=1$].}
	\label{FIG:Fixed-sample-IF-k100}
\end{figure}

Figure \ref{FIG:Fixed-sample-IF-k100} shows the fixed sample influence functions under the model assumption
for different types of tails with $k=100$ and different values of contamination direction $j_0$.
The boundedness of the MDPDEs with $\alpha>0$ are clear from the figure. 
However, the influence functions become flatter as we take contamination in more extreme observations.
Interestingly, note also that the influence functions at positive $\gamma$ (Pareto-Type tail)
and negative $\gamma $ (Weibull-Type tail) are almost symmetrically opposite 
to each other in nature with respect to the value $0$ and the Gumbel-Type tails with $\gamma=0$ 
have influence functions lying in between the above two.

Next consider the more general case of contamination in all the $Y_j$s and 
define the corresponding MDPDE of $\gamma$ as 
$\gamma_\epsilon = T_\alpha(G_{1,\epsilon}, \cdots, G_{j_0, \epsilon}, \cdots, G_{k-1, \epsilon})$, 
where $G_{j, \epsilon} = (1-\epsilon) G_{j} + \epsilon \wedge_{t_j}$ for all $j$. 
The contamination points in this case are  $\mathbf{t} = (t_1, \ldots, t_{k-1})$.
Then, we can derive the fixed-sample influence function 
of $T_\alpha$ at the true distribution proceeding as before.
For simplicity we only present the simplified results under Assumption (\ref{EQ:ERM_def1}) as given by 
\begin{eqnarray}
&& IF_k(\mathbf{t}; T_\alpha, \underline{\mathbf{F}}) =  \nonumber\\
&=& \frac{(1+\alpha)^3}{(1+\alpha^2)} 
\left[\sum_{j=1}^{k-1}\frac{1}{\theta_j^{\alpha-2}}\widetilde{J}\left(\frac{j}{k+1}\right)^2\right]^{-1} 
\sum_{j=1}^{k-1}~\widetilde{J}_\alpha\left(\frac{j}{k+1}\right)
\left[(t_j - \theta_{j})e^{-\frac{\alpha t_j}{\theta_{j}}} + \frac{\alpha\theta_{j}}{(1+\alpha)}\right].
\nonumber 
\end{eqnarray}
Note that, here also, the influence function of $T_\alpha$ is bounded for all $\alpha>0$ 
with any choice of $k$, but it is unbounded at $\alpha=0$. 
This again shows the robustness of the proposed MDPDE of tail index 
under contamination in all $Y_j$s over the existing non-robust MB estimator.

\begin{figure}[!th]
	\centering
	\subfigure[$\gamma=1$, $j_0 = k/2$]{
		\includegraphics[width=0.3\textwidth] {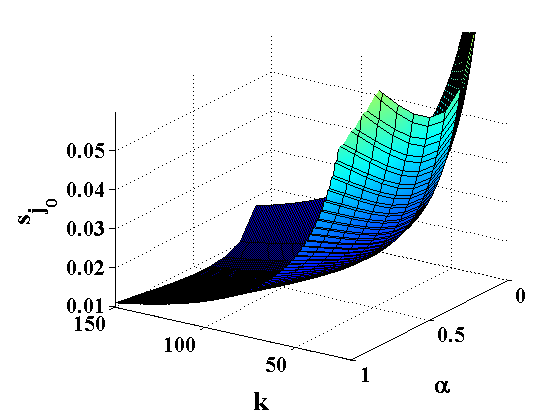}
		\label{FIG:Sens_g1p_J2}}
	~ 
	\subfigure[$\gamma=0$, $j_0 = k/2$]{
		\includegraphics[width=0.3\textwidth] {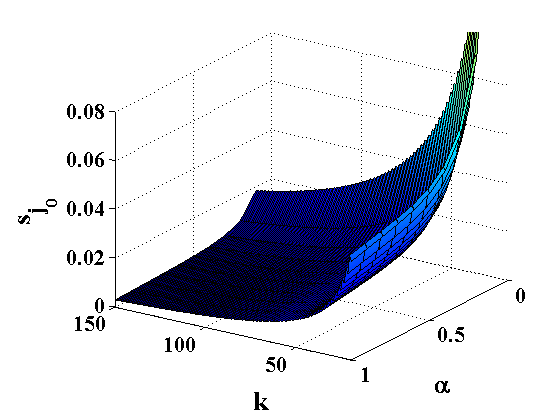}
		\label{FIG:Sens_g0_J2}}
	~ 
	\subfigure[$\gamma=-1$, $j_0 = k/2$]{
		\includegraphics[width=0.3\textwidth] {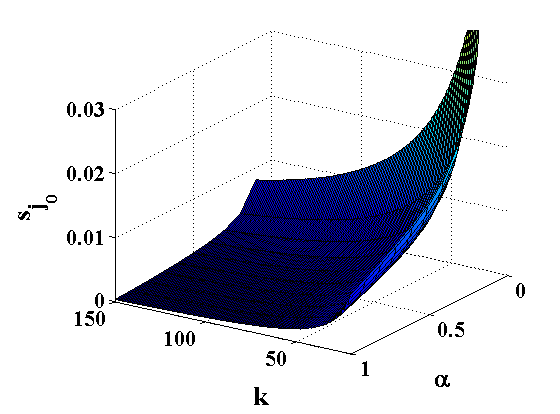}
		\label{FIG:Sens_g1n_J2}}
	\\ 
	\subfigure[$\gamma=1$, $j_0 = k/5$]{
		\includegraphics[width=0.3\textwidth] {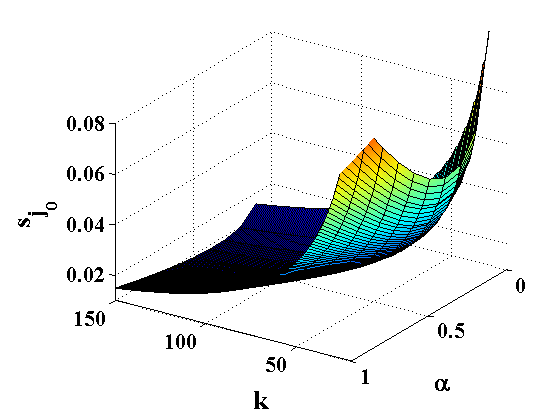}
		\label{FIG:Sens_g1p_J5}}
	~ 
	\subfigure[$\gamma=0$, $j_0 = k/5$]{
		\includegraphics[width=0.3\textwidth] {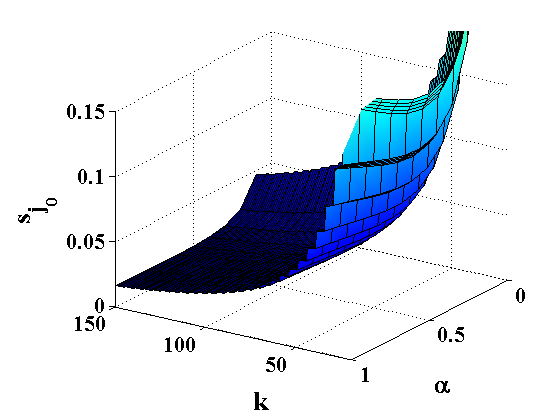}
		\label{FIG:Sens_g0_J5}}
	~ 
	\subfigure[$\gamma=-1$, $j_0 = k/5$]{
		\includegraphics[width=0.3\textwidth] {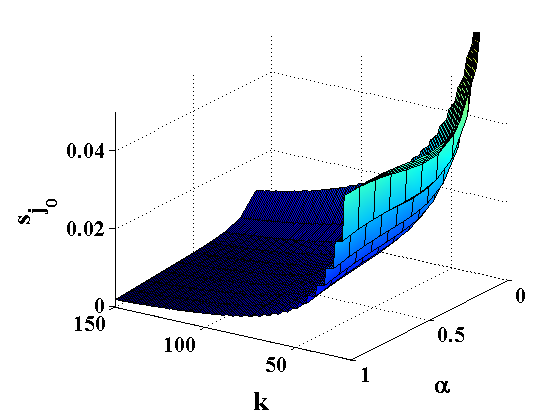}
		\label{FIG:Sens_g1n_J5}}
	\caption{Gross-error sensitivity $s_{j_0} (T_\alpha, \underline{\mathbf{F}})$ over the tuning parameters $k$ and $\alpha$ for different types of tails with contamination direction $j_0=k/2$ and $j_0=k/5$.}
	\label{FIG:Fixed-sample-Sensitivity}
\end{figure}

Next, in order to examine the effects of $k$ and $\alpha>0$ on the extent of robustness,
we consider the ``{\it Gross-Error Sensitivity}" measure (Hampel, 1968) defined as 
\begin{align}
& s(T_\alpha, \underline{\mathbf{G}}) = 
\sup_t \left\{||IF(t, T_\alpha, \underline{\mathbf{F}})||\right\}. 
\label{EQ:sensitivity_def}
\end{align}
As the influence function gives us the indication of asymptotic bias under contamination,
this measure will reflect the maximum possible value of the bias that 
a estimator may concede under infinitesimal contamination.
Thus, smaller the value of $s(T_\alpha, \underline{\mathbf{G}})$, higher the stability of 
the estimator $T_\alpha$ with respect to contamination, implying greater robustness.
In case of the proposed MDPDE of $\gamma$, 
the form of the gross-error sensitivity under contamination only in $Y_{j_0}$ is given by 
\begin{eqnarray}
&& s_{j_0} (T_\alpha, \underline{\mathbf{F}}) = 
\sup_{t_0} \left\{||IF_{k,j_0}(t_0; T_\alpha, \underline{\mathbf{F}})||^2\right\}^{\frac{1}{2}} \nonumber\\
&=& \left\{\begin{array}{l c l}
\frac{(1+\alpha)^2(e^{-(1+\alpha)} + \alpha^2)}{(1+\alpha^2)\alpha}  
\left[\sum_{j=1}^{k-1}\frac{1}{\theta_j^{\alpha-2}}\widetilde{J}\left(\frac{j}{k+1}\right)^2\right]^{-1}
\widetilde{J}\left(\frac{j_0}{k+1}\right) \frac{1}{\theta_{j_0}^{\alpha-1}} & \mbox{ if } & \alpha>0,\\
\infty  & \mbox{ if } & \alpha=0.
\end{array}\right.\nonumber
\end{eqnarray}
Figure \ref{FIG:Fixed-sample-Sensitivity} 
shows the values of this sensitivity measure $s_{j_0} (T_\alpha, \underline{\mathbf{F}})$ 
over the tuning parameters $k$ and $\alpha$ for different types of tail 
with contamination direction $j_0=k/2$ and $j_0=k/5$. 
Clearly the sensitivity measure $s_{j_0} (T_\alpha, \underline{\mathbf{F}})$ 
decreases as both the tuning parameter $\alpha$ and 
the number $k$ of extreme observations used increases;
it in fact tends to infinity as $\alpha, k \rightarrow 0.$
Further, the rate of change in the values of $s_{j_0} (T_\alpha, \underline{\mathbf{F}})$ 
with respect to $k$ is more in case of positive $\gamma$ (Pareto-type tail) compared 
to the case of negative $\gamma$ (Weibull-type tail) and 
the case with $\gamma=0$ (Gumbel type tail) has values in between these two. 
On the other hand, the  dependency of the sensitivity measure on $\alpha$ 
is more strict  for the Weibull-Type tails compared to the Pareto-type tails.
However, in all the cases, the choice $\alpha \geq 0.3$ and $k \geq 100$ 
gives quite small values of $s_{j_0} (T_\alpha, \underline{\mathbf{F}})$  
implying strong robustness properties of the corresponding MDPDEs. 
With respect to the contamination direction $j_0$, there is not much of a difference in the 
nature of sensitivity over the tuning parameter $\alpha$ and $k$; 
only its value increases slightly with $j_0$.
The sensitivity for contamination in more than one or in all the observations can be obtained similarly;
it has been seen to have exactly the same behavior as the case of contamination in one direction 
and hence those details are not presented here for brevity.

Note that the above fixed sample influence function and the sensitivity measure 
depend on sample size $n$ through the parameter $k$; in usual practice
we assume $k=\delta n $ for some small fraction $\delta$. 
Thus, it would be interesting for  a practitioner working with a large data set to know the 
similar robustness properties of the proposed estimator 
as $n\rightarrow\infty$. The asymptotic influence function obtained by taking the limit as $k\rightarrow\infty$
in the above fixed sample influence function provides us such asymptotic robustness analysis;
note that $k\rightarrow\infty$ as $n\rightarrow\infty$ by the usual assumption.
Also for any fixed $j_0$, 
$$
\lim\limits_{k\rightarrow\infty} ~ \theta_{j_0} = \left\{ 
\begin{array}{lcl}
\gamma, & \mbox{ if } & \gamma >0,\\
0, & \mbox{ if } & \gamma \leq 0,
\end{array}\right.
$$
and 
$$
\lim\limits_{k\rightarrow\infty} ~ \widetilde{J}\left(\frac{j_0}{k+1}\right) = \left\{ 
\begin{array}{lcl}
\gamma, & \mbox{ if } & \gamma >0,\\
0, & \mbox{ if } & \gamma \leq 0.
\end{array}\right.
$$
Using these, one can derive the asymptotic influence function under contamination only in  one fixed direction
as given by 
\begin{eqnarray}
IF_{j_0}(t_0; T_\alpha, \underline{\mathbf{G}}) 
&=& \lim\limits_{k\rightarrow\infty} ~ IF_{k,j_0}(t_0; T_\alpha, \underline{\mathbf{G}}) = 0.
\label{EQ:IF_asymp_MDPDE_cont1}
\end{eqnarray}
Thus all the MDPDEs including the maximum likelihood (or, the MB) estimator ($\alpha=0$) will 
be unaffected under contamination only in one fixed $Y_j$ 
provided we have a large enough sample size which is in-line with our intuition.
However, the most interesting case is the contamination in all $Y_j$s due to some heavy contamination in the original sample $X_i$s.
We will assume that the contamination points also go to infinity with the sample size, i.e., 
assume $t_j = \psi(t, j/(k+1))$ with $\psi$ being an positive unbounded function of $t$.
Then the asymptotic influence function can be seen to have the form
\begin{eqnarray}
&& IF(t_0; T_\alpha, \underline{\mathbf{G}}) 
= \lim\limits_{k\rightarrow\infty} ~ IF_{k}(t_0; T_\alpha, \underline{\mathbf{G}}) \nonumber\\
&=& \frac{(1+\alpha)^3}{(1+\alpha^2)}  
\frac{\int_0^1 ~ \widetilde{J}(u) \left[\left(\psi(t,u) - \frac{\gamma}{1-u^\gamma}\right)
e^{-\frac{\alpha}{\gamma}\psi(t,u)(1-u^\gamma)} + \frac{\alpha\gamma}{(1+\alpha)(1-u^\gamma)} \right]du}{
\gamma^{\alpha+2}\int_0^1 ~ \widetilde{J}(u)^2 (1-u^\gamma)^{\alpha-2}du}.
\nonumber 
\end{eqnarray}
Note that, this asymptotic influence function at $\alpha=0$ depends on the outlier parameter $t$ 
through a linear function of $\int_0^1 \psi(t,u) du$, which is unbounded in $t$ (since the positive integral is so).
However, the same for $\alpha>0$ depends on $t$ through an exponential function of $[-\int_0^1 \psi(t,u)du]$
and hence bounded which implies the robustness of the proposed MDPDEs.

\subsection{Some Comments on the Asymptotic Properties}

Although the main focus of the present paper is methodological, 
here we will provide some comments and give some indication about the asymptotic distribution of the proposed MDPDE.
For simplicity, let us first assume that the approximation (\ref{EQ:ERM_def1}) holds exactly. This is clearly a much stronger assumption compared to the usual second order assumptions of the extreme value theory. 
However, under this stronger assumption, the asymptotic properties of the proposed MDPDE will follow directly
from the results of Ghosh and Basu (2013) provided the required conditions (Assumptions (A1)--(A7) of their paper)
can be verified.
In this case our $j^{\rm th}$ model density of the transformed variable $Y_j$ is simply 
the exponential density with mean $\theta_j$ which yields the following simple result. 

\begin{theorem}
Consider the above mentioned set-up for estimation of the tail index $\gamma$ and assume that 
the exponential regression approximation (\ref{EQ:ERM_def1}) holds uniformly over the support of $Y_j$s
with $k\rightarrow\infty$ as $n\rightarrow\infty$.
Then there exists a consistent sequence $\gamma_{k,n}^{(\alpha)}$ of roots of the minimum density power divergence 
estimating equation (\ref{EQ:MDPDE_est_eqn}) with tuning parameter $\alpha$. Further, the asymptotic distribution of 
$\sqrt{k-1}~\left(\gamma_{k,n}^{(\alpha)} - \gamma\right)$  is normal with mean $0$ and 
variance $\sigma_\gamma^2/a_\gamma^2$, provided this asymptotic variance exits. 
Here, we have defined, for $\gamma \neq 0$ 
\begin{eqnarray}
a_\gamma &=& \frac{(1+\alpha^2)}{(1+\alpha)^3 \gamma^{\alpha+2}}
~\int_0^1(1-u^\gamma - \gamma u^\gamma\log u)^2 (1-u^\gamma)^{\alpha-2}du, \nonumber\\
\sigma_\gamma^2 &=& \left[\frac{(1+4\alpha^2)}{(1+2\alpha)^3} - \frac{\alpha^2}{(1+\alpha)^4}\right] 
\frac{1}{\gamma^{2\alpha+2}} ~\int_0^1(1-u^\gamma - \gamma u^\gamma\log u)^2 (1-u^\gamma)^{2\alpha-2}du.\nonumber
\end{eqnarray}
and for $\gamma=0$,
\begin{eqnarray}
a_0 = \frac{(1+\alpha^2)}{4(1+\alpha)^3}~\int_0^1 (-\log u)^{\alpha+2} du,
~\sigma_0^2 = \left[\frac{(1+4\alpha^2)}{(1+2\alpha)^3} - \frac{\alpha^2}{(1+\alpha)^4}\right] 
\frac{1}{4} ~\int_0^1 (1-\log u)^{2\alpha+2}.\nonumber
\end{eqnarray}
\label{THM:asymp_distr_est}
\end{theorem}
\noindent
\textbf{Proof:}
Since the exponential regression approximation (\ref{EQ:ERM_def1}) holds uniformly over the support 
of $Y_j$s, asymptotically we can work with the independent variables $W_j$, $j=1,\ldots,k-1$, 
where each $W_j$ follows an exponential distribution with mean $\theta_j$ and 
the required asymptotic distribution of the tail index estimator will be the same as the 
distribution of the minimum DPD estimator of $\gamma$ under this set-up.
Now, a simple but lengthy calculation (as presented in Appendix \ref{APP: Proof_asymp_est}) 
shows that Conditions (A1)--(A7) of Ghosh and Basu (2013) hold for this particular exponential regression model.
Then, a direct application of Theorem 3.1 of Ghosh and Basu (2013) proves the 
existence of a consistence sequence of estimators $\gamma_{k,n}^{(\alpha)}$ with 
$$
\Omega_k^{-1/2}\Psi_k\sqrt{k-1}(\gamma_{k,n}^{(\alpha)}-\gamma) \mathop{\rightarrow}^\mathcal{D} N(0,1),
$$
as $k\rightarrow\infty$ (or $n\rightarrow\infty$), where 
$$
\Omega_k = \left[\frac{(1+4\alpha^2)}{(1+2\alpha)^3} - \frac{\alpha^2}{(1+\alpha)^4}\right]  
~\frac{1}{k-1}\sum_{j=1}^{k-1} \widetilde{J}_\alpha\left(\frac{j}{k+1}\right)^2\theta_j^2,
$$
and
$$
\Psi_k = \frac{(1+\alpha^2)}{(1+\alpha)^3} ~\frac{1}{k-1}\sum_{j=1}^{k-1} \widetilde{J}\left(\frac{j}{k+1}\right) 
\theta_j^{\alpha-2},
$$
Thus the theorem follows by noting the fact that $\Omega_k \rightarrow \sigma_\gamma^2$ and 
$\Psi_k \rightarrow a_\gamma$ as $k \rightarrow \infty$.
\hfil{$\square$}


It follows from the theorem that the proposed MDPDEs are asymptotically unbiased and 
its asymptotic variance (obtained by a a simple numerical integration) increases slightly 
as $\alpha$ increases.

However, it is worthwhile to note here that we have proved the above theorem under a very strong condition 
that  may not hold for usual parametric models. 
So, it would be of great importance to derive the asymptotic distribution of the proposed estimator 
under a more general set-up as considered in Matthys and Beirlant (2003) for the likelihood case. 
Although we think that this could be done by extending the calculations of Matthys and Beirlant (2003)
for the case of general MDPDE (as pointed out by a referee also), we do not have a concrete proof at this time. 
We hope to pursue these theoretical arguments in our subsequent research work and 
focus more on the method and its applications through an extensive simulation study as presented in the next section.

\section{Numerical Illustrations}
\label{SE:Simulation}

\subsection{Models and Simulation Set-Up}

We will now present an extensive simulation study to illustrate the performance of the proposed estimators $\hat{\gamma}_{ER,k}^{(\alpha)}$. 
We consider several classes of distributions having different types of tail following Matthys and Beirlant (2003).
Specifically, we consider three model having positive tail index, 
two having zero tail index and two having negative tail index as described below:
\begin{enumerate}
\item[(M1)]	Student's t-distribution with degrees of freedom $\nu$ which has a positive tail index given by 
$\gamma=1/\nu$.
\item[(M2)]	 Burr($\beta, \tau, \lambda$) distribution defined by the survival function 
$\bar{F}(x) = \left(1+\frac{x^\tau}{\beta}\right)^{-\lambda}$. It also has a positive tail index given by 
$\gamma=1/\tau\lambda$.
\item[(M3)]	 Fr\'{e}chet($\gamma$) distribution with tail index $\gamma>0$ whose distribution function is given by 
${F}(x) = \exp\left(-x^{-1/\gamma}\right)$. 
\item[(M4)]	Standard log-normal distribution having tail index $\gamma=0$.
\item[(M5)]	Weibull($\lambda, \tau$) distribution defined by the survival function 
$\bar{F}(x) = \exp\left(-\lambda{x^\tau}\right)$. It also has zero tail index.
In our simulation, we have taken $\lambda=1$ and $\tau=2$.
\item[(M6)]	Uniform($0,1$) distribution having right end-point $1$ and negative tail index $\gamma=-1$.
\item[(M7)]	 Reversed Burr($\beta, \tau, \lambda$) distribution defined by the survival function 
$\bar{F}(x) = \left(1+\frac{(x_{+} - x)^\tau}{\beta}\right)^{-\lambda}$. 
It has a finite right end-point $x_{+}$ and negative tail index given by $\gamma=-1/\tau\lambda$.
We have taken $x_{+}=2$ in our simulation.
\end{enumerate}

Using these models and their combinations, 
we will create several interesting scenarios with and without contamination and 
examine the proposed estimator in terms of both the bias and the MSE.
Under each scenario, we simulate samples of size $n=500$ and 
estimate the tail index using our proposal and some existing methods. 
Based on $100$ replications, we compute the empirical estimate of the bias and the MSE of 
the proposed estimators $\hat{\gamma}_{ER,k}^{(\alpha)}$ for different values of $\alpha$;
note that the case $\alpha=0$ gives the MB estimator which is expected to be 
non-robust but has smaller bias under pure data.
In the next subsections, we will present the results for some interesting scenarios.

As we have noted earlier, there are only a few robust estimators of the tail index available 
in the literature for $\gamma>0$. So, for the cases $\gamma \leq 0$, 
we can only compare the proposed estimators with the existing non-robust estimators, e.g., the MB estimator. 
For the cases with $\gamma>0$, we have considered the non-robust MB and Hill estimators 
as a point of reference along with the existing robust methods of Vandewalle et al.~(2007) and Kim and Lee (2008). 
The estimator proposed by Kim and Lee (2008), to be denoted by $\hat{\gamma}_{KL,k}^{(\alpha)}$,  
is defined as the minimizer of the DPD with tuning parameter $\alpha$ under the assumption of 
exponentiality of the log-relative excess; it also includes the proposal of Vandewalle et al.~(2007)
as a special case at $\alpha=1$. 
Another existing proposal of estimating Pareto-type tail, given by Vandewalle et al.~(2004),
makes use of a robust regression method (Marazzi and Yohai, 2004) under 
a similar exponential regression model as in our work
but involves several complications related to the deepest  regression method used 
(as noted in Appendix \ref{APP: Vandewalle_2004_est});
so we have not considered their proposal in our simulation study.

\subsection{Performance of $\hat{\gamma}_{ER,k}^{(\alpha)}$ under pure data}

We first consider the performance of $\hat{\gamma}_{ER,k}^{(\alpha)}$  under pure data with no contamination.
We have performed the simulation study for all the models (M1)--(M7) as described previously
and plot their empirical absolute biases and MSEs in Figures \ref{FIG:Sim_Cont_0_M1}--\ref{FIG:Sim_Cont_0_M7} respectively. 
For the first three models having positive $\gamma$, 
the same empirical measures are also presented for the estimators $\hat{\gamma}_{KL,k}^{(\alpha)}$ of Kim and Lee (2008).

Note that, when the data comes from the pure t-distribution (Figure \ref{FIG:Sim_Cont_0_M1}), 
the Hill's estimator has the minimum possible MSE but relatively high bias for lower values of $k$.
The MB estimators with moderately large $k$ give significantly less bias with competitive MSE. 
The estimators $\hat{\gamma}_{KL,k}^{(\alpha)}$ and $\hat{\gamma}_{ER,k}^{(\alpha)}$ with positive $\alpha$ perform 
similar to the Hill's estimator and the MB estimator respectively with slightly more bias and MSE.
In particular, for any $\alpha$, our proposal $\hat{\gamma}_{ER,k}^{(\alpha)}$ performs 
significantly better than $\hat{\gamma}_{KL,k}^{(\alpha)}$ in terms of bias and also has competitive MSE.
Similar observations can also be made in the case of a Burr distribution with $(\beta,~\tau,~\lambda) = (1,~1,~1)$ and $\gamma=1$
(Figure \ref{FIG:Sim_Cont_0_M2}) or a Fr\'{e}chet distribution with $\gamma=0.5$ (Figure \ref{FIG:Sim_Cont_0_M3}). 

For the Gumbel type tails (Figure \ref{FIG:Sim_Cont_0_M4}--\ref{FIG:Sim_Cont_0_M5}) also, 
the proposed estimators $\hat{\gamma}_{ER,k}^{(\alpha)}$ give quite 
satisfactory results at some moderate value of $k$, where its bias and MSE are very close to 
those of the MB estimators for all $\alpha>0$.

Similar results have also been obtained for the Weibull-type tails 
as presented in Figure \ref{FIG:Sim_Cont_0_M6}--\ref{FIG:Sim_Cont_0_M7}.

\subsection{Performance of $\hat{\gamma}_{ER,k}^{(\alpha)}$ under contamination by the same distribution family but having a different tail index}

%

Now let us consider the robustness aspect of the proposed estimators
by studying their performance under contaminated samples.
In this subsection, we assume that the actual observations and the contaminated observations
are both coming from the same distribution family but having different tail index.

First we take the samples from a t-distribution with $\nu=2$ 
(so that $\gamma=0.5$) and contaminate a certain percentage of each sample 
by observations from another t-distribution with $\nu=1/3$ ($\gamma=3$). 
The corresponding values of the empirical biases and MSEs of $\hat{\gamma}_{ER,k}^{(\alpha)}$  and 
$\hat{\gamma}_{KL,k}^{(\alpha)}$ are shown in Figures \ref{FIG:Sim_Cont_5_t1} and \ref{FIG:Sim_Cont_15_t1} respectively 
for $5\%$ and $15\%$ contaminations.
Clearly the robust estimators  $\hat{\gamma}_{KL,k}^{(\alpha)}$ and $\hat{\gamma}_{ER,k}^{(\alpha)}$ with a positive $\alpha$ 
perform much better than the Hill's estimator and the MB estimator respectively. 
Further, our proposed estimator $\hat{\gamma}_{ER,k}^{(\alpha)}$  with a large $k$ yields 
lower bias and MSE compared to the existing robust estimator of Kim and Lee (2001).
 
Figures \ref{FIG:Sim_Cont_5_F1} and \ref{FIG:Sim_Cont_15_F1} present the results for a 
similar set-up with the Fr\'{e}chet distribution having $\gamma=0.5$ for original data and $\gamma=3$ for the contaminated observations.
Once again the proposed estimator  $\hat{\gamma}_{ER,k}^{(\alpha)}$ with a positive $\alpha$ and a large $k$ 
performs better than the MB estimator and also the existing robust estimators $\hat{\gamma}_{KL,k}^{(\alpha)}$.

%

The performances of $\hat{\gamma}_{ER,k}^{(\alpha)}$ under a similar scenario with Burr distribution is also seen 
to be competitive with the existing $\hat{\gamma}_{KL,k}^{(\alpha)}$ 
(Figures \ref{FIG:Sim_Cont_5_Burr1}--\ref{FIG:Sim_Cont_15_Burr1}).


\subsection{Performance of $\hat{\gamma}_{ER,k}^{(\alpha)}$ under contaminations by a different distribution having the same tail type}

Let us now consider a more general case where we consider data from a t-distribution 
with $\nu=2$ ($\gamma=0.5$) and contaminate a certain percentage of each sample 
by observations from a Fr\'{e}chet distribution having $\gamma=3$.
The resulting biases and MSEs are shown in Figures \ref{FIG:Sim_Cont_5_t2} and \ref{FIG:Sim_Cont_15_t2} respectively
for $5\%$ and $15\%$ contaminations. 
Again our proposal $\hat{\gamma}_{ER,k}^{(\alpha)}$ with positive $\alpha$ performs 
much better compared to the non-robust Hill estimator and the MB estimator 
and is also competitive to the robust estimators $\hat{\gamma}_{KL,k}^{(\alpha)}$.
%


Next we consider the Gumbel type tails and take the samples from a standard log-normal distribution with 
contamination from standard Weibull distribution. 
From the results presented in Figures \ref{FIG:Sim_Cont_05_LN1} and \ref{FIG:Sim_Cont_15_LN1}, 
it is clear that the estimator $\hat{\gamma}_{ER,k}^{(\alpha)}$ with a moderate positive $\alpha$ and a large $k$ 
has similar MSE but improved bias compared to the MB estimator.

\subsection{Performance of $\hat{\gamma}_{ER,k}^{(\alpha)}$ under contaminations by a different distribution from a different tail type}

In this subsection, we consider the cases where contaminated part of the data comes from a completely different distribution
of different tail type. Form our extensive simulation study considering several combinations of models (M1)--(M7),
we only report the results of the following interesting cases in Figures \ref{FIG:Sim_Cont_05_LN2}--\ref{FIG:Sim_Cont_15_RBurr2}
respectively:
\begin{enumerate}
\item[(i)] Sample is from the model (M4) having $\gamma=0$ and the contamination is from the model (M1) having $\nu=1/3$. 

\item[(ii)] Sample is from the model (M5) having $\gamma=0$ and the contamination is from the model (M1) having $\nu=1/3$.

\item[(iii)] Sample is from the model (M5) having $\gamma=0$ and the contamination is from the model (M6) having $\gamma=-1$.

\item[(iv)] Sample is from the model (M6) having $\gamma=-1$ and the contamination is from the model (M5) having $\gamma=0$.

\item[(v)] Sample is from the model (M6) having $\gamma=-1$ and the contamination is from the model (M1) having $\nu=1/3$. 

\item[(vi)] Sample is from the model (M7) having $(\beta,~\tau,~\lambda) = (1,~1,~1)$ so that $\gamma=-1$  
and the contamination is from the model (M2) having $(\beta,~\tau,~\lambda) = (4,~0.25,~1)$ and $\gamma=1$.
\end{enumerate}

In all the cases, the proposed estimator $\hat{\gamma}_{ER,k}^{(\alpha)}$ with a positive $\alpha$ provides 
an improvement in terms of both the bias and the MSE compared to the existing non-robust MB estimators
implying their high stability in the presence of contamination in the data.


\section{On the choice of tuning parameters $k$ and $\alpha$}
\label{SEC:choice_parameter}

As we have seen in our extensive simulation study, the performance of the estimators $\hat{\gamma}_{ER,k}^{(\alpha)}$ 
differ significantly for different values of the tuning parameters $\alpha$ and $k$ for all the three tail types. 
So, we need to choose these parameters carefully 
in order to obtain the optimum results both in terms of robustness and efficiency.
Based on the findings of previous sections, we note the followings in respect to the 
dependence of $\hat{\gamma}_{ER,k}^{(\alpha)}$ on $\alpha$ and $k$:

\begin{enumerate}
\item When there is no contamination in data, the estimators $\hat{\gamma}_{ER,k}^{(\alpha)}$ have least
MSE at $\alpha=0$ for any fixed $k$ (it is in fact the MB estimators); MSE increases slightly 
as $\alpha$ increases.

\item Under the pure model, if we choose a large enough $k$ then  
the performances of the estimators $\hat{\gamma}_{ER,k}^{(\alpha)}$ having different values of 
small positive $\alpha$ do not differ significantly.

\item For the contaminated samples, the estimators $\hat{\gamma}_{ER,k}^{(\alpha)}$ with positive $\alpha$ 
perform much better than the case $\alpha=0$ (MB estimator); however, their performances are mostly similar
for all $\alpha \geq 0.3$.

\item When there is contamination in the data with a Pareto type tail, 
our proposed estimators $\hat{\gamma}_{ER,k}^{(\alpha)}$ with positive $\alpha$ perform 
slightly better or at competitive level with the robust estimators of Kim and Lee (2008).
However, the estimators of Kim and Lee generate optimum bias and MSE at a smaller value of $k$,
whereas our proposed estimators give minimum bias and MSE at a larger value of $k$ for any fixed $\alpha>0$.

\item For most of the cases with contaminated samples from Gumbel or Weibull type tails also,
the estimators $\hat{\gamma}_{ER,k}^{(\alpha)}$ with any fixed $\alpha>0$ generate optimum bias and MSE 
at a large value of $k$.

\item For all the three types of tails, the bias and MSE of $\hat{\gamma}_{ER,k}^{(\alpha)}$ at any fixed $\alpha >0$
decreases as $k$ increases; however, beyond a moderately large value near $k=2n/5$ or $k=n/2$ (we have $n=500$ in our simulation)
the rate of change becomes quite small.  
\end{enumerate}

It is clear from the above observations that
the tuning parameter $\alpha$ controls the trade-off between efficiency and robustness of the estimators 
$\hat{\gamma}_{ER,k}^{(\alpha)}$ of tail index. 
A similar effect of the tuning parameter $\alpha$ 
is observed in other minimum DPD estimators also; 
see, e.g., Basu et al.~(1998) and Ghosh and Basu (2013).
Thus, our empirical suggestion for the choice of $\alpha$ in any practical case is $\alpha\approx 0.3$
generating only a slight loss in efficiency with much better performance with respect to contamination. 

On the other hand the parameter $k$ affects mainly the bias of the estimators $\hat{\gamma}_{ER,k}^{(\alpha)}$
and as $k$ increases both bias and MSE decreases. This effect of $k$ on the performance of  
$\hat{\gamma}_{ER,k}^{(\alpha)}$ possibly comes from the violation of our main assumption (\ref{EQ:ERM_def1})
at smaller values of $k$; as $k$ increases the assumption of exponential regression model gives 
a better approximation to the model. Further, beyond $k=2n/5$ or $k=n/2$ (we have $n=500$ in our simulation), these model approximation 
is good enough so that there is not much of a improvement in bias (and MSE) of the estimators as seen in 
the simulation study. Therefore, we suggest based on our empirical findings, we suggest that $k$ in between $2n/5$ to $n/2$ 
will be a reasonable choice while applying the proposed methodology to any practical situation.

\section{Bias Correction and Real Data Examples}
\label{SEC:Bias_corr}

Note that the bias of the proposed estimators, 
although smaller compared to the existing methods in most of the cases, comes mainly from the violation of 
the  exponential model Assumption (\ref{EQ:ERM_def1}) through the choice of $k$. 
So, there is a scope of improving the proposal through suitable bias correction 
while applying it to any real life dataset. 
Here we will present one possible approach for bias correction using a refinement of  Assumption (\ref{EQ:ERM_def1}) 
as done for the MB estimator in Matthys and Beirlant (2003).

In order to refine Assumption (\ref{EQ:ERM_def1}), let us consider the notations and assumptions of Section \ref{SEC:MLE_ERM}.
As proved in Section 4 of Matthys and Beirlant (2003), for any $\gamma\in \mathbb{R}$, Assumption (\ref{EQ:EVT_def2}) implies 
the existence of a slowly varying function $l$ and a function $d$ with $\pm d$ slowly varying and 
$d(t)\rightarrow\gamma$  as $t\rightarrow\infty$, such that for all $t, \lambda >0$,
\begin{eqnarray}
 \frac{Q(\lambda t) - Q(t)}{a_Q(t)}  = \frac{1}{d(t)}
 \left(t^\gamma\frac{l(\lambda t)}{l(t)} - 1\right).
\label{EQ:EVT_def3}
\end{eqnarray} 
Following Matthys and Beirlant (2003), we can get a bias-corrected estimate of the tail index $\gamma$
by assuming a second order condition as follows. We assume the existence of a real constant $\rho \leq 0$
and a rate function $b$ satisfying $b(t)\rightarrow 0$ as $t\rightarrow\infty$ such that for all $\lambda \geq 1$,
\begin{eqnarray}
\lim\limits_{t\rightarrow\infty}\frac{1}{b(t)}\log \left(\frac{l(\lambda t)}{l(t)}\right) = 
\left\{\begin{array}{l c l}
\frac{\lambda^\rho-1}{\rho} & \mbox{ for } & \rho \ne 0,\\
\log \lambda & \mbox{ for } & \rho = 0,
\end{array}\right.,
\label{EQ:EVT_def4}
\end{eqnarray} 
Then, following an argument  similar to the derivation of (\ref{EQ:ERM_def1}), we can get a refined model approximation as
(see Section 4.1 of Matthys and Beirlant, 2003, for details of the derivation)
\begin{eqnarray}
j\log\left(\frac{X_{(n-j+1)} - X_{(n-k)}}{X_{(n-j)} - X_{(n-k)}}\right)
&\mathop{\sim}^d&  \frac{\gamma + \beta \left(\frac{j}{k+1}\right)^{-\rho}}{1 - \left(\frac{j}{k+1}\right)^\gamma
	\exp\left(\beta\frac{\left(\frac{j}{k+1}\right)^{-\rho}-1}{-\rho}\right)} E_{k-j+1}, ~~~~
\label{EQ:ERM_def5} 
\end{eqnarray}
for $j = 1, \ldots, k-1$. This refined approximation reduces to the model approximation (\ref{EQ:ERM_def1}) whenever $\beta=0$.

Now, we can easily obtain the simultaneous minimum DPD estimator of the parameters $\gamma$, $\beta$ and $\rho$
following the method proposed in Section \ref{SEC:MDPDE_ERM} with the new parametrization 
$$
\theta_j =  \frac{\gamma + \beta \left(\frac{j}{k+1}\right)^{-\rho}}{1 - \left(\frac{j}{k+1}\right)^\gamma
	\exp\left(\beta\frac{\left(\frac{j}{k+1}\right)^{-\rho}-1}{-\rho}\right)}
$$
based on the refined approximation (\ref{EQ:ERM_def5}). We can easily compute the estimators 
numerically and the robustness properties can be derived similarly as before. 
In particular at $\alpha=0$, it will coincide with the bias-corrected MB (MB$_c$, say) estimator of Matthys and Beirlant (2003, Section 4.1).
In the following, we will examine the performance of this bias-correction method for the MDPDEs
of the tail index $\gamma$ through some interesting real data examples.

\begin{figure}[h]
	\centering
	\subfigure[Newcomb Data (histogram)]{
		\includegraphics[width=0.45\textwidth, height=0.35\textwidth] {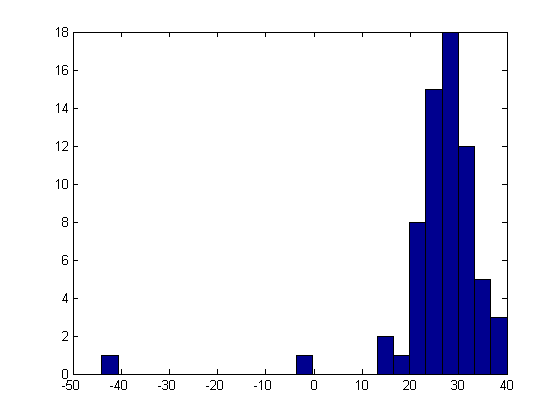}
		\label{FIG:ex1}}
	~ 
	\subfigure[Danish Data]{
		\includegraphics[width=0.45\textwidth, height=0.35\textwidth] {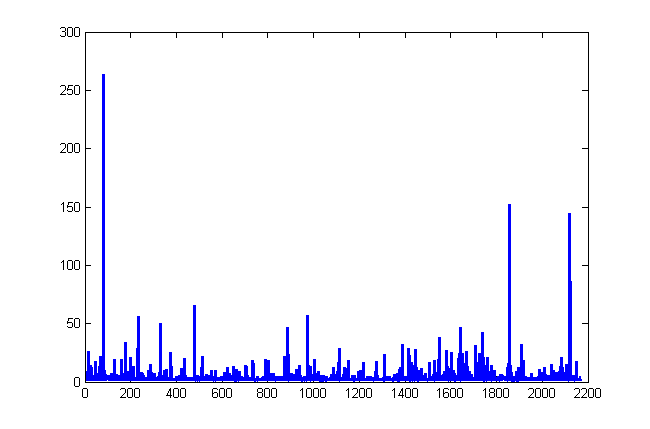}
		\label{FIG:ex2}}
	\caption{Plots of the real data examples}
	\label{FIG:ex12}
\end{figure}

\bigskip\noindent
\textbf{Example 1: (Newcomb data)}\\
Our first example involves Newcomb's light speed data (Stigler, 1977, Table 5)
which have been analyzed by several authors in the context of robust inference. 
There are two large (negative) outliers among the 66 data points (Figure \ref{FIG:ex1}); 
the normal model provides an excellent fit to these data if one ignores the two outliers. 
So, in the absence of the outliers, these data have a tail measure $0$;
but the outliers falsely indicate its heavy tail (like a $t$-distribution) 
having a positive tail index if the usual  non-robust estimators of tail index are used.
Note that, since the original data have zero tail index, the existing robust estimator of  Kim and Lee (2008) cannot be used
but our proposed methodology can be applied. Table \ref{TAB:ex1_est} presents the estimated tail indices using our proposed 
methodology with bias correction for different values of $\alpha$ and $k=15$. 
Although the MB$_c$ is non-robust yielding positive tail index, 
the robust nature of our proposal at $\alpha>0.1$ is clear from the results.

\begin{table}[!th]
	\caption{\label{TAB:ex1_est} Bias corrected $\hat{\gamma}_{ER,k}^{(\alpha)}$ with $k=15$ for the Newcomb data}
	\centering
	\begin{tabular}{r| r r r r r r r } \hline
$\alpha$ & 0 (MB$_c$) &	0.1	& 0.2 & 0.4	& 0.5	& 0.6 & 1\\\hline
$\hat{\gamma}$ & 0.5474	& $-0.0065$ & 	$0.0000$ & 	4.60E-11	& 7.35E-11	& 4.62E-24	& 1.36E-25\\	
\hline
	\end{tabular}
\end{table}

\bigskip\noindent
\textbf{Example 2: (Danish data)}\\
Now we will consider the Danish dataset consisting of 2492 large fire insurance claims 
in Denmark during the period 3rd January, 1980 to 31st December, 1990,
obtained from the R package {\it `evir'}. This dataset has been used by Pak (2013) to illustrate the 
usefulness of the minimum density power divergence estimator for the composite lognormal-Pareto model.
Here we will estimate the tail index for these data; clearly the data have a heavy tail with positive tail index. 
It has been plotted in Figure \ref{FIG:ex2}; note that all the data-points are smaller than 66 except 
for the three large values of 263.25, 152.41 and 144.66. 
In order to examine the robustness of our proposal, we contaminate these data in two different ways;  
scheme (i) replaces the highest observation (263.25) by a much larger value of 10000 and 
scheme (ii) replaces all the three indicated large observations by the value 70. 
The estimated tail indices, bias corrected $\hat{\gamma}_{ER,k}^{(\alpha)}$, for all the three cases 
have been shown in Table \ref{TAB:ex2_est}; we have taken $k=950$ 
which is near the choice $k=2n/5$ as suggested in the previous section.

Note that, under scheme (i), one additional large observation makes the non-robust MB$_c$ estimator larger 
compared to the value under the original data. On the other hand, contamination in scheme (ii) 
can be thought of as contamination from the Weibull class of distribution having finite 
right end-point (at $70$) and negative tail index; 
hence the non-robust MB$_c$ estimator yields a negative tail measure.
However, the proposed estimators $\hat{\gamma}_{ER,k}^{(\alpha)}$ (bias-corrected) 
with moderately large positive $\alpha$ lead to stable inference under both kinds of contamination
indicating their strong robustness properties.

\begin{table}[!th]
	\caption{\label{TAB:ex2_est} Bias corrected $\hat{\gamma}_{ER,k}^{(\alpha)}$ with $k=950$ for the Danish data}
	\centering
	\begin{tabular}{r| r r r r r r r r} \hline
		$\alpha$ & 0 (MB$_c$) &	0.1	& 0.2 & 0.3& 0.4	& 0.5	& 0.6 & 1\\\hline
		Original data & 0.50	& 0.66	& 0.72	& 0.78 &	0.78	& 0.77	& 0.76	&	0.69\\	
		Scheme (i) & 0.83 & 	0.82 &	0.81	& 0.80 &	0.79	& 0.78	& 0.76	&	0.70 \\	
		Scheme (ii) & $-1.62$	& $-1.64$	& 0.62	& 0.65 & 0.70	& 0.73 & 	0.72	&	0.67\\	
		\hline
	\end{tabular}
\end{table}

\bigskip\noindent
\textbf{Example 3: (Austrian EUSILC data, 2006)}\\
In this example we will apply our proposed method of robust tail index estimation to individual income data. 
Such data can be modeled by a Pareto distribution with the parameter being the reciprocal of the tail index (see Alfons et al. 2013b). 
Data from the European Union statistics on income and living conditions (EUSILC) annual panel survey conducted in 
European Union member states and other European countries provide a basis for computing several economic indicators 
in order to measure risk of poverty and social exclusion in Europe. There could be several possible outliers in the survey data
that influence the estimated economic indicators and hence affect the overall social decisions based on these indicators. 
Austrian and Belgian EUSILC income survey data from 2005 to 2006 had been used by Alfons et al. (2013b) to study the effect of 
outliers on the Gini coefficients of household income which illustrated the usefulness of robust methodologies for these indicators. 
The study of several other indicators based on these data is available in Hulliger et al. (2011). 
Since many economic indicators like income, expenditure etc. follow heavy-tailed distributions, 
robust estimators of their tail indices are extremely important.  

Due to the confidentiality of the actual survey data, here we have used 
a close-to-reality simulated dataset, as in Section 7 of Alfons et al.~(2013b), 
which was kindly provided by Professor Peter Filzmoser (personal communication). 
This dataset contains the net income of 9519 individuals from 4641 households, generated synthetically 
from Austrian EUSILC survey data from 2006 using the methodology proposed in Alfons et al.~(2011a) 
through the R package ``{\it simPopulation}" (Alfons and Kraft, 2012). 
A similar synthetic dataset has also been used in the vignette ``{\it leaken-pareto}" (Alfons et al., 2011b) and 
implemented in the R package ``{\it laeken}" (Alfons and Templ, 2013; Alfons et al., 2013a).

\begin{figure}[h]
	\centering
	\subfigure[Original Data]{
		\includegraphics[width=0.47\textwidth] {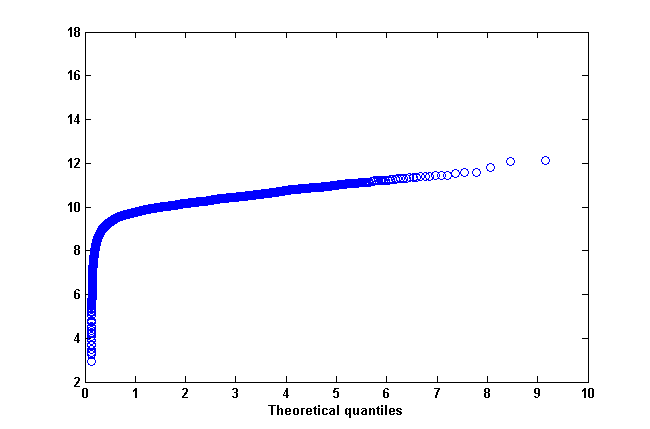}
		\label{FIG:ex3a}}
	~ 
	\subfigure[Data with One Outlier]{
		\includegraphics[width=0.47\textwidth] {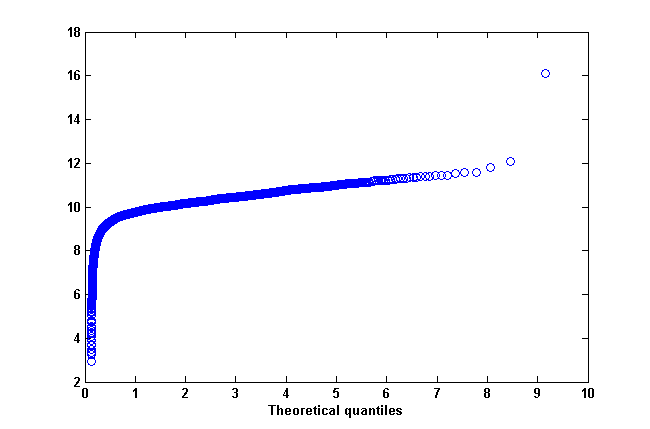}
		\label{FIG:ex3o}}
	\caption{Pareto-Quantile plots for EUSILC Data}
	\label{FIG:ex3}
\end{figure}

Since the dataset was simulated with full control, there are no outliers in the original data and it can be well fitted 
by a Pareto distribution; see Figure \ref{FIG:ex3a} for the Pareto-Quantile plot (Beirlant et al., 1996) 
of the data whose slope gives the indication of its tail index. 
In order to examine the robustness of our proposal, we replace the largest income by $10000000$ 
and estimate the tail index of this contaminated dataset; the corresponding Pareto-Quantile plot is shown in Fig \ref{FIG:ex3o}.
The classical Hill's tail-index estimator for Pareto modeling (with threshold $k=100$) changes 
from $0.2315$ to $0.2714$ due to the insertion of this single  outlier.
Table \ref{TAB:ex3_est} provides the bias corrected estimators $\hat{\gamma}_{ER,k}^{(\alpha)}$  for several values of $\alpha$
for the contaminated data; here we have taken $k=1500$.  
Robustness of our proposals with $\alpha>0.1$ is again evident from the estimates obtained 
which are much closer to the original value of 0.23.

\begin{table}[!th]
	\caption{\label{TAB:ex3_est} Bias corrected $\hat{\gamma}_{ER,k}^{(\alpha)}$ with $k=1500$ for the EUSILC data with one artificial outlier}
	\centering
	\begin{tabular}{r| r r r r r r r r} \hline
		$\alpha$ & 0 (MB$_c$) &	0.1	& 0.2 & 0.3& 0.4	& 0.5	& 0.6 & 1\\\hline
Estimates & 0.30	& 0.26	& 0.24	& 0.23	& 0.21	& 0.21	& 0.21 &	0.23 \\	
		\hline
	\end{tabular}
\end{table}


\section{Conclusion}\label{SEC:cnclusion}

The present paper considers the problem of estimating the tail index under data contamination 
and proposes a robust estimator based on the density power divergence and an exponential model 
approximation to the true data which works equally well in all the three types of tail.
Thus, it needs no prior information on the type of tail and ensures that
 a ``better" estimator compared to the  existing ones would be obtained when some 
outlying observations are present in the sample. 
In this present paper we have given more emphasis on this robustness aspect of the proposed estimators
rather than its asymptotic properties and the illustrations have been provided through a 
theoretical analysis based on influence functions as well as through an extensive simulation study and real data examples.
We have also provided some indication about its asymptotic efficiency under different tail-types. 
Based on the findings of our extensive simulation study, we have also 
made an empirical suggestion about the choice of tuning parameters so that any researcher
can apply this proposal to any real life data set.
More detailed theoretical properties of the proposed estimators will be considered in our future research work.

\begin{acknowledgements}
The author would like to thank Prof. Ayanendranath Basu  of the Indian Statistical Institute, India, 
for his valuable comments about this work and Prof. Peter Filzmoser  of Vienna University of Technology, Austria
for kindly providing the dataset used in Example 3 of Section \ref{SEC:Bias_corr}.
The author also wishes to thank two anonymous referees for their remarks
that have led to an improved version of the paper.
\end{acknowledgements}


\bigskip
\appendix
\section{Assumptions (A1)--(A7) of Ghosh and Basu (2013) under the Assumed Exponential Regression Model}
\label{APP: Proof_asymp_est}
We assume the set-up of an exponential regression model, 
where the random variables $W_1, \ldots, W_n, \ldots$ are  independent, but for each $j$, 
$W_j$ follows an exponential distribution with mean 
$\theta_j =  \frac{\gamma }{1 - \left(\frac{j}{k+1}\right)^\gamma}$. 
In this paper we have approximated the distribution of the transformed variable $Y_j$,
as defined in Section \ref{SEC:MLE_ERM}, by the distribution of $W_j$ for each $j=1, \ldots, k-1$.
Now we will present a brief argument to show that Assumptions (A1)--(A7) of 
Ghosh and Basu (2013), required for asymptotic consistency and normality of the MDPDE,  hold 
under the present set-up of exponential regression model.

First note that Assumption (A1)--(A3) and (A5) hold directly from the form of an exponential distribution function.
Next, as shown in the proof of Theorem \ref{THM:asymp_distr_est}, the matrix $J^{(i)}$,
as per the notation of Ghosh and Basu (2013), is a positive scalar given by 
$$
J^{(i)} = \frac{(1+\alpha^2)}{(1+\alpha)^3} \widetilde{J}\left(\frac{i}{k+1}\right) 
\theta_i^{\alpha-2}.
$$
So, the matrix $\Psi_n$ is in fact a positive scalar with 
$\lambda_0 = \lim\limits_{k\rightarrow\infty}\Psi_n = a_\gamma >0$;
this implies that (A4) also holds.
Finally we need to prove three limiting statements of Assumptions (A6) and (A7) of Ghosh and Basu (2013).
We only present the proof of first one, namely 
(noting that we are dealing with scalar parameter $\gamma$ here)
\begin{eqnarray}
\lim_{N\rightarrow\infty} \sup_{k>1} \left\{ \frac{1}{k-1} \sum_{i=1}^{k-1}
E \left[|\nabla_{g}V_i(W_i;\theta)| I(|\nabla_{g}V_i(W_i;\theta)| > N)\right]\right\} = 0.
\label{EQ:GB13_A6_1}
\end{eqnarray}
Here $\nabla_g$ represents the derivative with respect to our parameter of interest $\gamma$.
The proof of the others are similar and hence omitted.

To prove (\ref{EQ:GB13_A6_1}), note that under this present model, we have for each $i$,
$$
\nabla_{g}V_i(W_i;\theta) = C_i \left[\frac{\alpha}{(1+\alpha)^2} 
+ \left(\frac{W_i}{\theta_i} - 1\right)e^{-\frac{\alpha W_i}{\theta_i}}\right] 
= C_i \psi\left(\frac{W_i}{\theta_i}\right),
$$
where $C_i= (1+\alpha)\widetilde{J}_\alpha\left(\frac{i}{k+1}\right) \theta_i$ and 
$\psi(w) = \frac{\alpha}{(1+\alpha)^2} + (w-1) e^{-\alpha w}$.
However, letting $W_i^* = \frac{W_i}{\theta_i}$, we get that $W_1^*, \ldots, W_{k-1}^*$ 
are independent and identically distributed observations from a standard exponential distribution with mean $1$.
So, we have
\begin{eqnarray}
&&  \frac{1}{k-1} \sum_{i=1}^{k-1}
E \left[|\nabla_{g}V_i(W_i;\theta)| I(|\nabla_{g}V_i(W_i;\theta)| > N)\right] \nonumber\\
&=&  \frac{1}{k-1} \sum_{i=1}^{k-1}
E \left[|C_i| \left|\psi\left(\frac{W_i}{\theta_i}\right)\right| 
I(|C_i| \left|\psi\left(\frac{W_i}{\theta_i}\right)\right|  > N)\right] \nonumber\\
&=&  \frac{1}{k-1} \sum_{i=1}^{k-1} |C_i| E\left[\left|\psi(W_i^*)\right| 
I( \left|\psi(W_i^*)\right|  > \frac{N}{\max_{1\leq i \leq k-1}|C_i|})\right] \nonumber\\
&=& E\left[ \left|\psi(W_1^*)\right| I( \left|\psi(W_1^*)\right|>\frac{N}{\max_{1\leq i \leq k-1}|C_i|})\right]
 \left(\frac{1}{k-1} \sum_{i=1}^{k-1}|C_i|\right). \nonumber
\end{eqnarray}
However, it is easy to check that both the terms $\left(\frac{1}{k-1} \sum_{i=1}^{k-1}|C_i|\right)$ and
$\left(\max_{1\leq i \leq k-1}|C_i|\right)$ are bounded as $k\rightarrow\infty$. 
Thus, by Dominated Convergence Theorem, we have 
$$
\lim\limits_{N\rightarrow\infty} ~ 
E\left[ \left|\psi_1(W_1^*)\right| I( \left|\psi_1(W_1^*)\right|>\frac{N}{\max_{1\leq i \leq k-1}|C_i|})\right]=0,
$$
and hence (\ref{EQ:GB13_A6_1}) holds.

\section{Some Comments on the Estimator Proposed by Vandewalle et al.~(2004)}
\label{APP: Vandewalle_2004_est}

Vandewalle et al.~(2004) presented an interesting and practically important footstep in statistics 
by justifying the necessity of combining two apparently contradictory theory of 
extreme value statistics and robust statistics, primarily for the Pareto-Type tails 
($\gamma>0$). They have used the robust regression method proposed by Marazzi and Yohai (2004)
and the exponential regression model developed in Beirlant et al.~(1999) given by 
\begin{equation}
\label{EQ:App2_eq1}
Y_j \sim_d \left(\gamma + b_{n,k}\left(\frac{\gamma}{k+1}\right)^{-\rho}\right) g_j, ~~~ j=1, \ldots, k,
\end{equation}
where $g_j$ are independent and identically distributed standard exponential random variables. 
The proposed estimator was examined through an interesting real data example 
where its robustness was illustrated clearly.

While developing the robust estimator, Vandewalle et al.~(2004) transformed the above model 
into a liner form given by Equation (3.1) of their paper, which reads
\begin{equation}
\label{EQ:App2_eq2}
Y_j \sim_d \gamma + b_{n,k}\left(\frac{\gamma}{k+1}\right)^{-\rho} + \gamma e_j, ~~~ j=1, \ldots, k,
\end{equation}
where $e_j = g_j -1$. Here comes our first little doubt by noting that the RHS of the 
Equations (\ref{EQ:App2_eq1}) and (\ref{EQ:App2_eq2}) are not equal; the closest form to 
the second that equals the first is 
$$
\gamma + b_{n,k}\left(\frac{\gamma}{k+1}\right)^{-\rho}g_j + \gamma e_j.
$$
So, it needs to be clarified the reason of dropping $g_j$ from the second term.
After assuming the linearized form (\ref{EQ:App2_eq2}), they have re-parametrize it as
\begin{equation}
\label{EQ:App2_eq3}
Y_j = \theta_i + \theta_2t_j + \sigma e_j,  ~~~ j=1, \ldots, k,
\end{equation}
where $ t_j = \left(\frac{\gamma}{k+1}\right)^{-\rho}$, $\theta_1 = \gamma$, $\theta_2= b_{n,k}$
and $\sigma =\gamma$. Then, for the case $\gamma > 0$, they have used
the robust regression method proposed by Marazzi and Yohai (2004) to estimate 
the parameters $(\theta_1, ~\theta_2, ~\sigma)$. This regression method  
has high breakdown and efficiency for usual regression set-up that they have noted 
for proposing the robust estimator of $\gamma$; However, the approach is computationally complicated. 
Moreover, under the transformed set-up  (\ref{EQ:App2_eq3}) it is to be noted that $\theta_1 = \sigma$;
this constraint needs to be taken care of while solving for the estimator numerically and 
 may have potential effect on the properties of the resulting estimator. 
This needs to be examined extensively through simulation or theoretical results,
that was missing in the work of Vandewalle et al.~(2004). 
They have also noted similar limitation of the work and made a comment in the ``conclusion" that 
they would consider this issues in their future work. 
Considering all this doubts, we have decided not to consider this proposal in our simulation studies.


\clearpage\newpage
\begin{figure}[f]
\centering
\includegraphics[width= 0.4\textwidth,height= 0.2\textwidth] {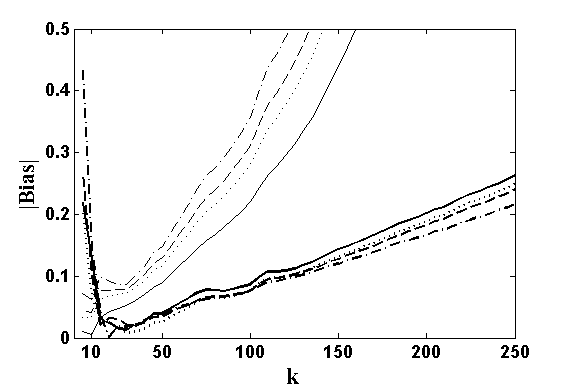}
\includegraphics[width= 0.4\textwidth,height= 0.2\textwidth] {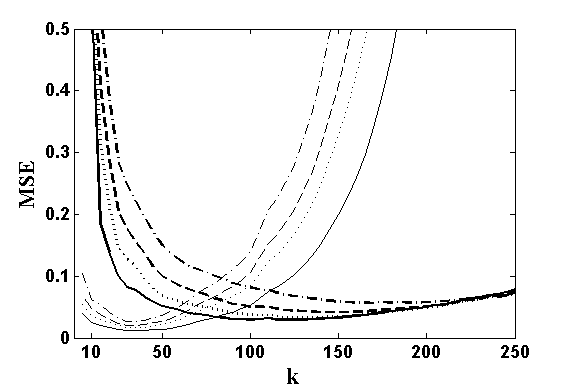}
\caption{The empirical bias and MSE of the estimators $\hat{\gamma}_{ER,k}^{(\alpha)}$ (thick lines) 
and $\hat{\gamma}_{KL,k}^{(\alpha)}$  (thin lines) for model (M1) having $\nu=2$ with no contamination 
[Solid line: $\alpha=0$, Dotted line: $\alpha=0.3$, Dashed-dotted  line: $\alpha=0.5$, Dashed line: $\alpha=1$].}
\label{FIG:Sim_Cont_0_M1}
\end{figure}

\begin{figure}[f]
\centering
\includegraphics[width= 0.4\textwidth,height= 0.2\textwidth] {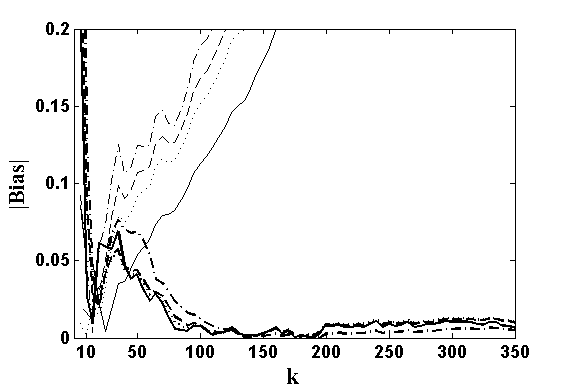}
\includegraphics[width= 0.4\textwidth,height= 0.2\textwidth] {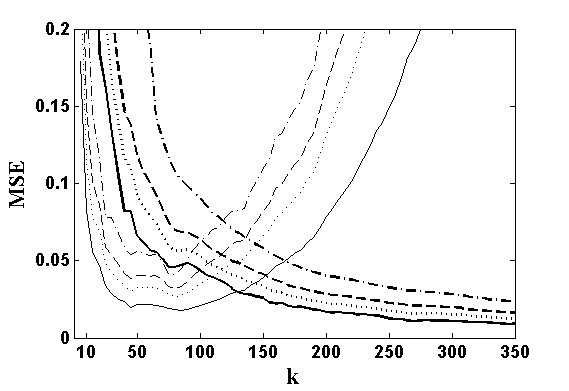}
\caption{The empirical bias and MSE of the estimators $\hat{\gamma}_{ER,k}^{(\alpha)}$ (thick lines) 
and $\hat{\gamma}_{KL,k}^{(\alpha)}$  (thin lines) for model (M2) having $(\beta,~\tau,~\lambda) = (1,~1,~1)$ 
with no contamination 
[Solid line: $\alpha=0$, Dotted line: $\alpha=0.3$, Dashed-dotted  line: $\alpha=0.5$, Dashed line: $\alpha=1$].}
\label{FIG:Sim_Cont_0_M2}
\end{figure}

\begin{figure}[f]
\centering
\includegraphics[width= 0.4\textwidth,height= 0.2\textwidth] {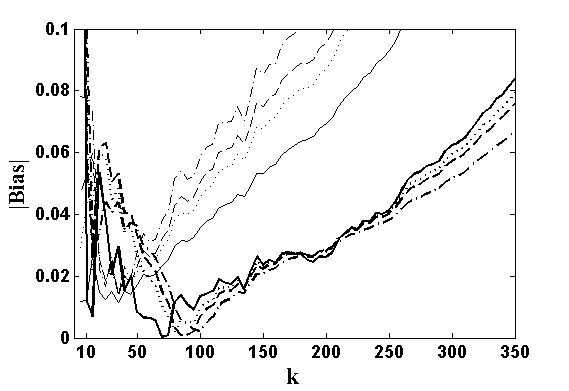}
\includegraphics[width= 0.4\textwidth,height= 0.2\textwidth] {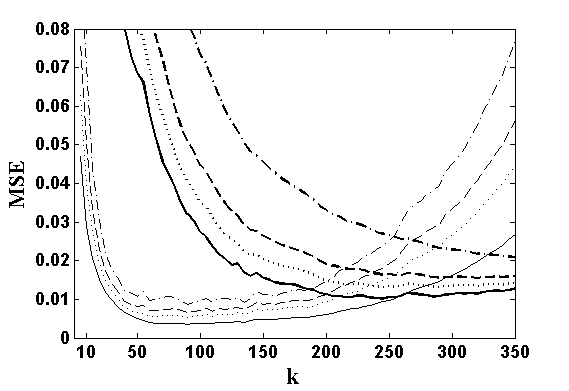}
\caption{The empirical bias and MSE of the estimators $\hat{\gamma}_{ER,k}^{(\alpha)}$ (thick lines) 
and $\hat{\gamma}_{KL,k}^{(\alpha)}$  (thin lines) for model (M3) having $\gamma=0.5$ with no contamination 
[Solid line: $\alpha=0$, Dotted line: $\alpha=0.3$, Dashed-dotted  line: $\alpha=0.5$, Dashed line: $\alpha=1$].}
\label{FIG:Sim_Cont_0_M3}
\end{figure}

\begin{figure}[f]
\centering
\includegraphics[width= 0.4\textwidth,height= 0.2\textwidth] {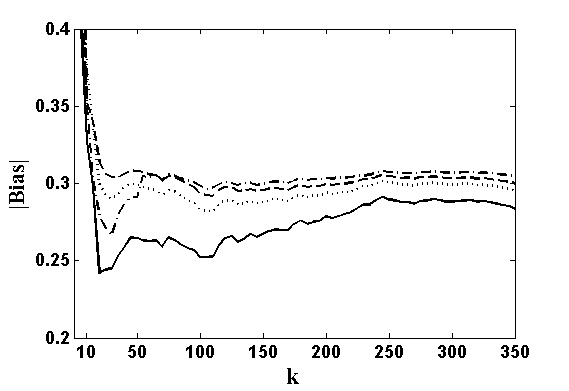}
\includegraphics[width= 0.4\textwidth,height= 0.2\textwidth] {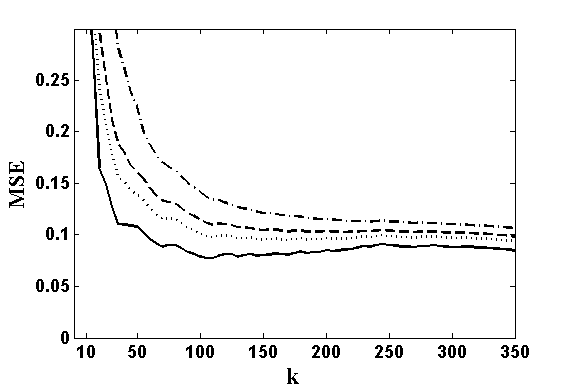}
\caption{The empirical bias and MSE of the estimators $\hat{\gamma}_{ER,k}^{(\alpha)}$ 
 for model (M4) having $\gamma=0$ with no contamination 
[Solid line: $\alpha=0$, Dotted line: $\alpha=0.3$, Dashed-dotted  line: $\alpha=0.5$, Dashed line: $\alpha=1$].}
\label{FIG:Sim_Cont_0_M4}
\end{figure}

\clearpage\newpage
\begin{figure}[f]
\centering
\includegraphics[width= 0.4\textwidth,height= 0.2\textwidth] {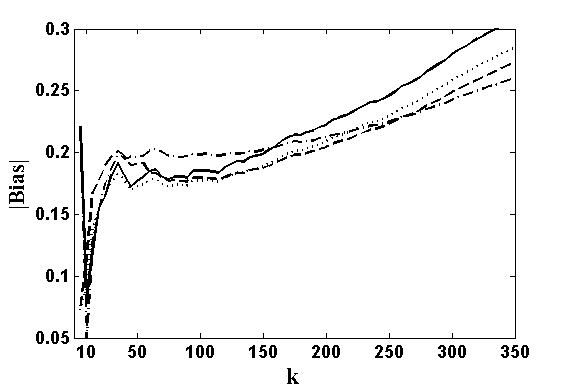}
\includegraphics[width= 0.4\textwidth,height= 0.2\textwidth] {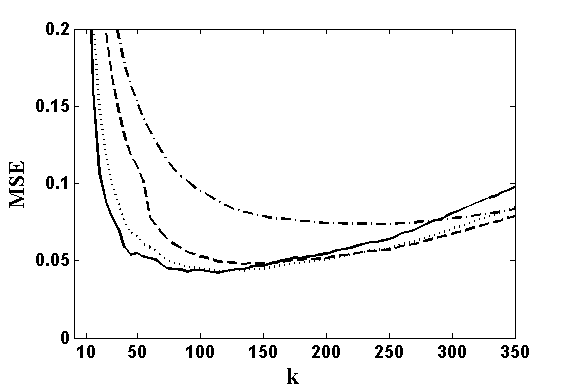}
\caption{The empirical bias and MSE of the estimators $\hat{\gamma}_{ER,k}^{(\alpha)}$ 
 for model (M5) having $\gamma=0$ with no contamination 
[Solid line: $\alpha=0$, Dotted line: $\alpha=0.3$, Dashed-dotted  line: $\alpha=0.5$, Dashed line: $\alpha=1$].}
\label{FIG:Sim_Cont_0_M5}
\end{figure}

\begin{figure}[f]
\centering
\includegraphics[width= 0.4\textwidth,height= 0.2\textwidth] {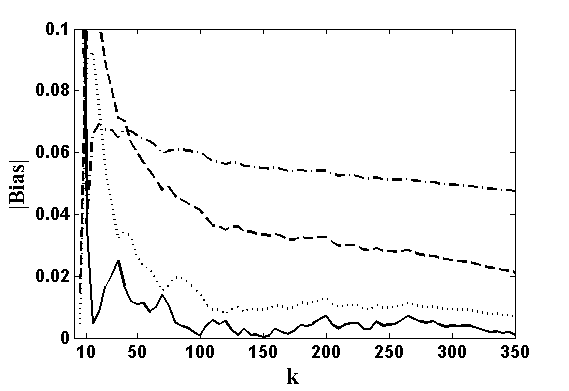}
\includegraphics[width= 0.4\textwidth,height= 0.2\textwidth] {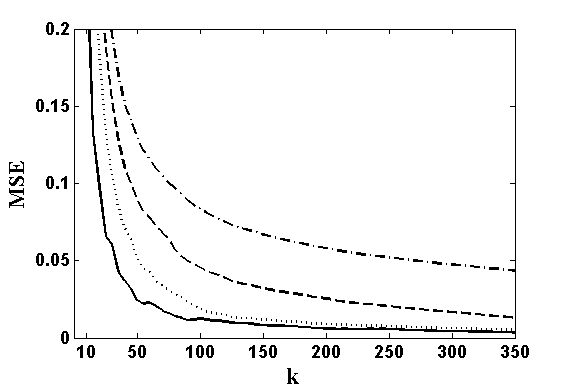}
\caption{The empirical bias and MSE of the estimators $\hat{\gamma}_{ER,k}^{(\alpha)}$
 for model (M6) having $\gamma=-1$ with no contamination 
[Solid line: $\alpha=0$, Dotted line: $\alpha=0.3$, Dashed-dotted  line: $\alpha=0.5$, Dashed line: $\alpha=1$].}
\label{FIG:Sim_Cont_0_M6}
\end{figure}

\begin{figure}[f]
\centering
\includegraphics[width= 0.4\textwidth,height= 0.2\textwidth] {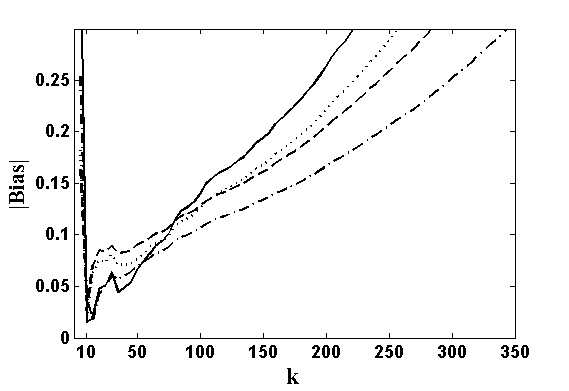}
\includegraphics[width= 0.4\textwidth,height= 0.2\textwidth] {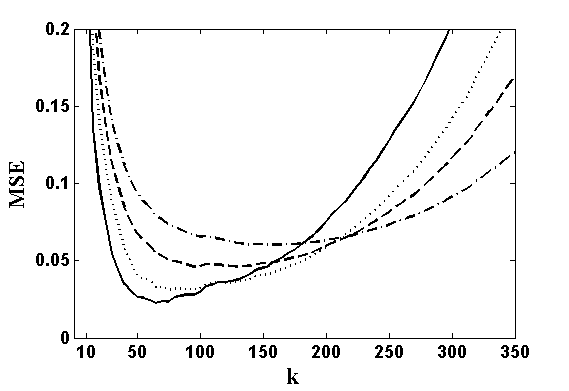}
\caption{The empirical bias and MSE of the estimators $\hat{\gamma}_{ER,k}^{(\alpha)}$ 
 for model (M7) having $(\beta,~\tau,~\lambda) = (1,~1,~1)$  with no contamination 
[Solid line: $\alpha=0$, Dotted line: $\alpha=0.3$, Dashed-dotted  line: $\alpha=0.5$, Dashed line: $\alpha=1$].}
\label{FIG:Sim_Cont_0_M7}
\end{figure}

\clearpage\newpage
\begin{figure}[f]
\centering
\subfigure[$5\%$ contamination]{
\includegraphics[width= 0.4\textwidth,height= 0.2\textwidth] {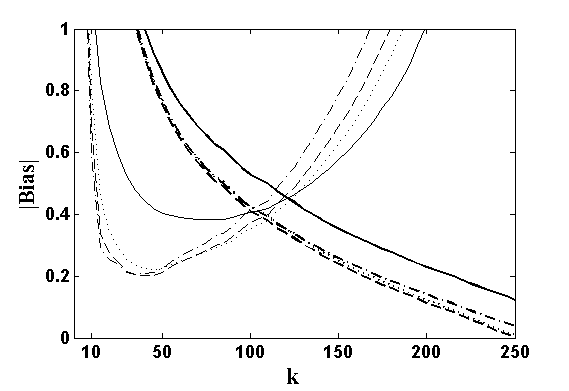}
\includegraphics[width= 0.4\textwidth,height= 0.2\textwidth] {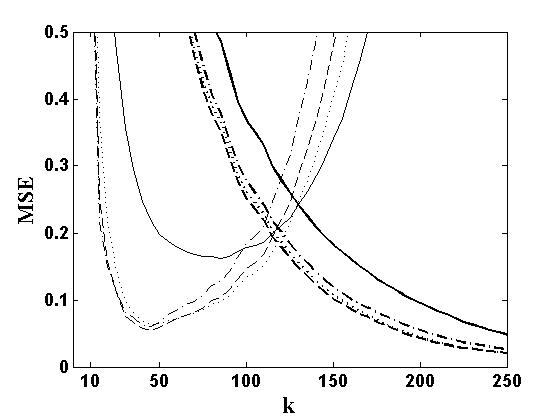}
\label{FIG:Sim_Cont_5_t1}}
\\
\subfigure[$15\%$ contamination]{
\includegraphics[width= 0.4\textwidth,height= 0.2\textwidth] {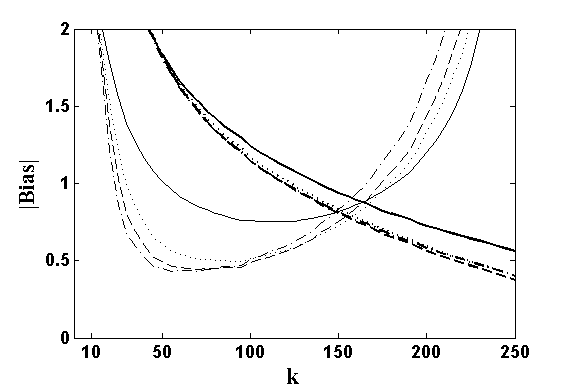}
\includegraphics[width= 0.4\textwidth,height= 0.2\textwidth] {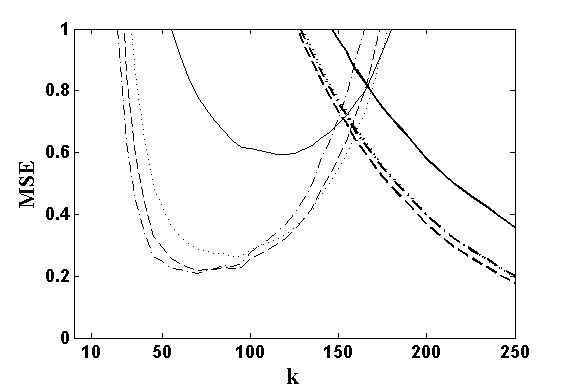}
\label{FIG:Sim_Cont_15_t1}}
\caption{The empirical bias and MSE of the estimators $\hat{\gamma}_{ER,k}^{(\alpha)}$ (thick lines) 
and $\hat{\gamma}_{KL,k}^{(\alpha)}$  (thin lines) for model (M1) having $\nu=2$  with contamination
by the same model (M1) having $\nu=1/3$  
[Solid line: $\alpha=0$, Dotted line: $\alpha=0.3$, Dashed-dotted  line: $\alpha=0.5$, Dashed line: $\alpha=1$].}
\end{figure}

\begin{figure}[f]
\centering
\subfigure[$5\%$ contamination]{
\includegraphics[width= 0.4\textwidth,height= 0.2\textwidth] {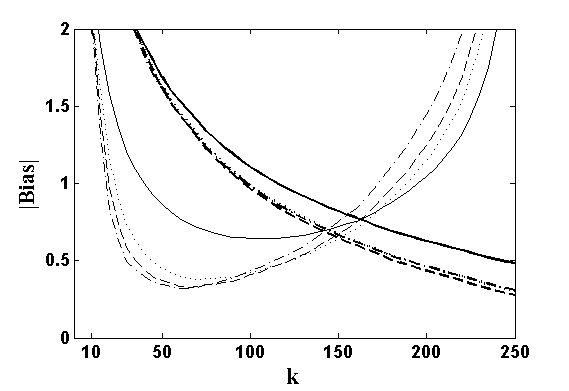}
\includegraphics[width= 0.4\textwidth,height= 0.2\textwidth] {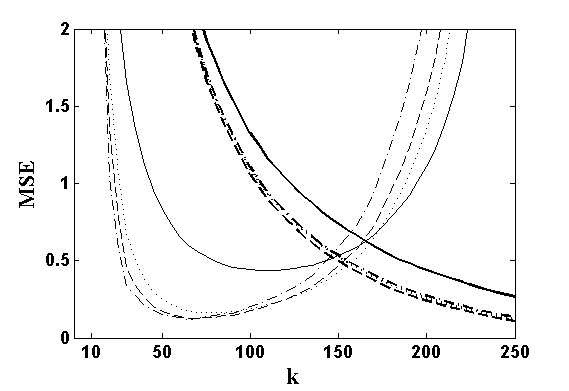}
\label{FIG:Sim_Cont_5_F1}}
\\
\subfigure[$15\%$ contamination]{
\includegraphics[width= 0.4\textwidth,height= 0.2\textwidth] {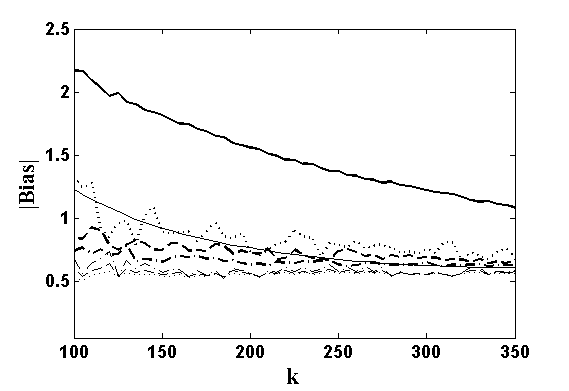}
\includegraphics[width= 0.4\textwidth,height= 0.2\textwidth] {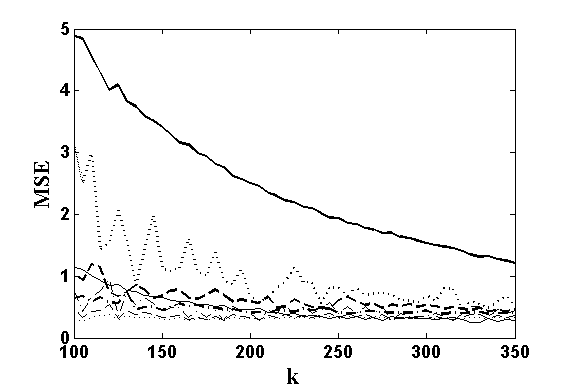}
\label{FIG:Sim_Cont_15_F1}}
\caption{The empirical bias and MSE of the estimators $\hat{\gamma}_{ER,k}^{(\alpha)}$ (thick lines) 
and $\hat{\gamma}_{KL,k}^{(\alpha)}$  (thin lines) for model (M3) having $\gamma=0.5$  with contamination
by the same model (M3) having $\gamma=3$ 
[Solid line: $\alpha=0$, Dotted line: $\alpha=0.3$, Dashed-dotted  line: $\alpha=0.5$, Dashed line: $\alpha=1$].}
\end{figure}

\clearpage\newpage
\begin{figure}[f]
\centering
\subfigure[$5\%$ contamination]{
\includegraphics[width= 0.4\textwidth,height= 0.2\textwidth] {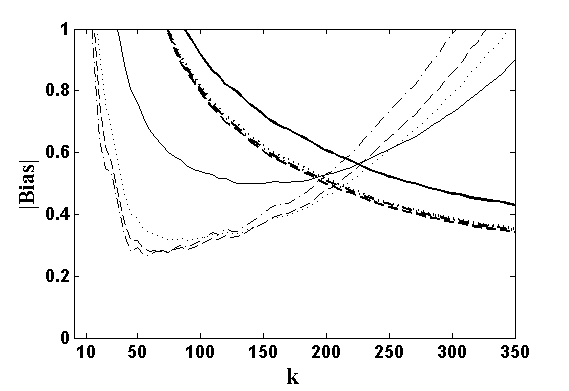}
\includegraphics[width= 0.4\textwidth,height= 0.2\textwidth] {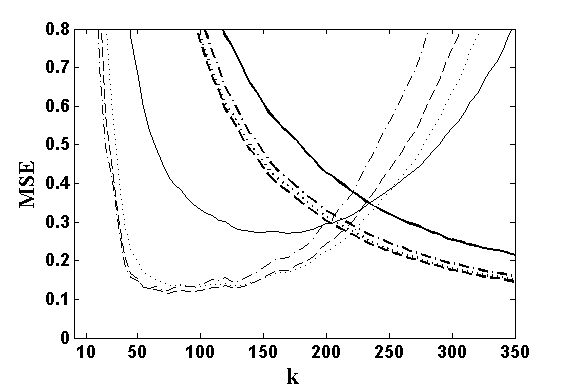}
\label{FIG:Sim_Cont_5_Burr1}}
\\
\subfigure[$15\%$ contamination]{
\includegraphics[width= 0.4\textwidth,height= 0.2\textwidth] {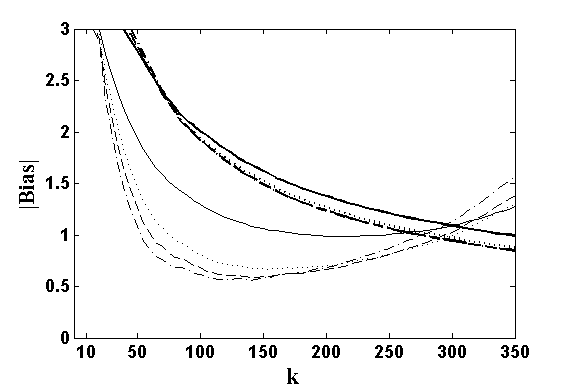}
\includegraphics[width= 0.4\textwidth,height= 0.2\textwidth] {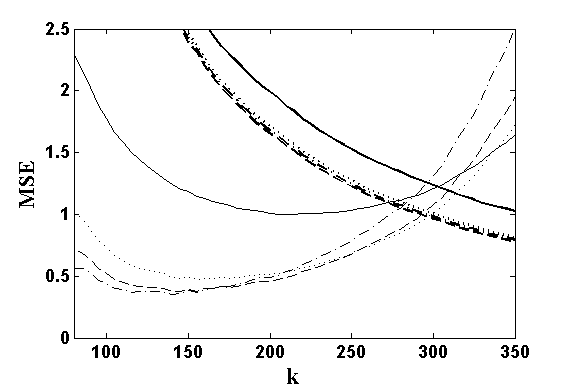}
\label{FIG:Sim_Cont_15_Burr1}}
\caption{The empirical bias and MSE of the estimators $\hat{\gamma}_{ER,k}^{(\alpha)}$ (thick lines) 
and $\hat{\gamma}_{KL,k}^{(\alpha)}$  (thin lines) for model  (M2) having $(\beta,~\tau,~\lambda) = (1,~1,~1)$  
with contamination by the same model (M2) having $(\beta,~\tau,~\lambda) = (1,~0.25,~1)$   
[Solid line: $\alpha=0$, Dotted line: $\alpha=0.3$, Dashed-dotted  line: $\alpha=0.5$, Dashed line: $\alpha=1$].}
\end{figure}

\begin{figure}[f]
\centering
\subfigure[$5\%$ contamination]{
\includegraphics[width= 0.4\textwidth,height= 0.2\textwidth] {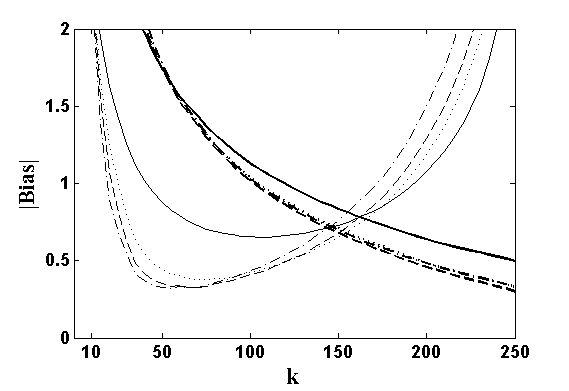}
\includegraphics[width= 0.4\textwidth,height= 0.2\textwidth] {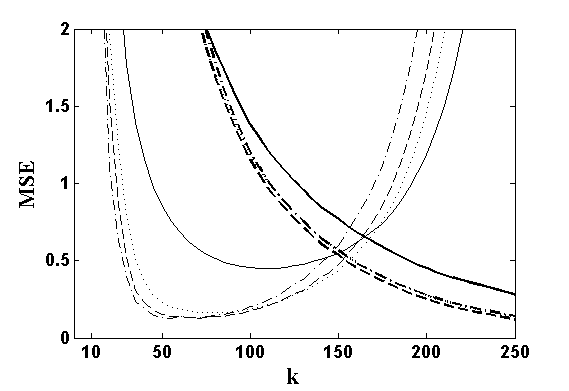}
\label{FIG:Sim_Cont_5_t2}}
\\
\subfigure[$15\%$ contamination]{
\includegraphics[width= 0.4\textwidth,height= 0.2\textwidth] {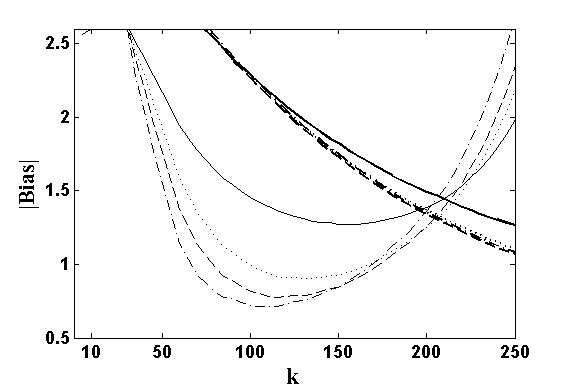}
\includegraphics[width= 0.4\textwidth,height= 0.2\textwidth] {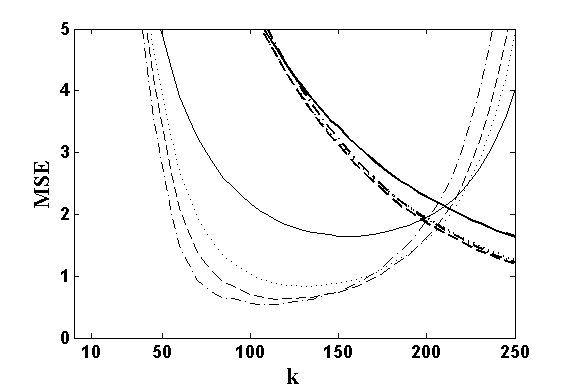}
\label{FIG:Sim_Cont_15_t2}}
\caption{The empirical bias and MSE of the estimators $\hat{\gamma}_{ER,k}^{(\alpha)}$ (thick lines) 
and $\hat{\gamma}_{KL,k}^{(\alpha)}$  (thin lines) for model (M1) having $\nu=2$  with contamination
by the model (M3) having $\gamma=3$ 
[Solid line: $\alpha=0$, Dotted line: $\alpha=0.3$, Dashed-dotted  line: $\alpha=0.5$, Dashed line: $\alpha=1$].}
\end{figure}

\begin{figure}[f]
\centering
\subfigure[$5\%$ contamination]{
\includegraphics[width= 0.4\textwidth,height= 0.2\textwidth] {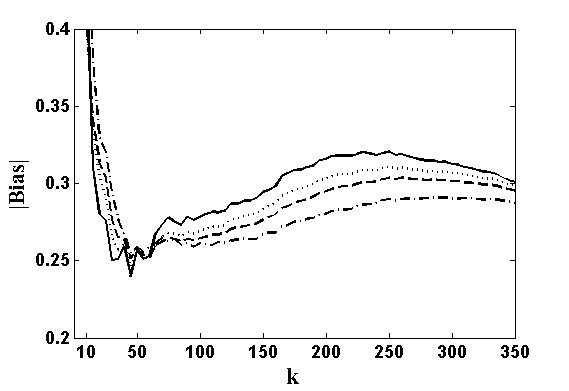}
\includegraphics[width= 0.4\textwidth,height= 0.2\textwidth] {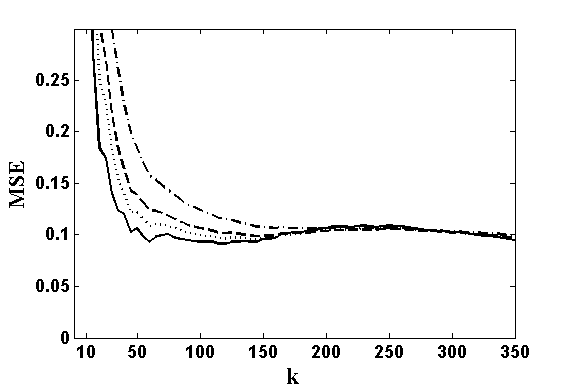}
\label{FIG:Sim_Cont_05_LN1}}
\\
\subfigure[$15\%$ contamination]{
\includegraphics[width= 0.4\textwidth,height= 0.2\textwidth] {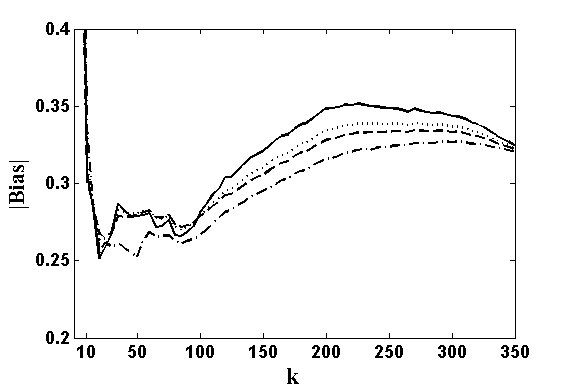}
\includegraphics[width= 0.4\textwidth,height= 0.2\textwidth] {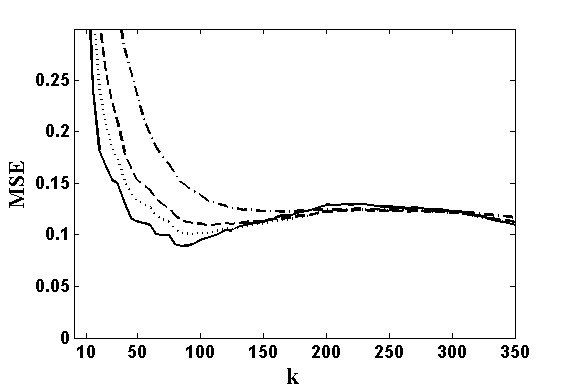}
\label{FIG:Sim_Cont_15_LN1}}
\caption{The empirical bias and MSE of the estimators $\hat{\gamma}_{ER,k}^{(\alpha)}$ 
 for model (M4) with contamination by the model (M5)  both having $\gamma=0$
[Solid line: $\alpha=0$, Dotted line: $\alpha=0.3$, Dashed-dotted  line: $\alpha=0.5$, Dashed line: $\alpha=1$].}
\end{figure}

\begin{figure}[f]
\centering
\subfigure[$5\%$ contamination]{
\includegraphics[width= 0.4\textwidth,height= 0.2\textwidth] {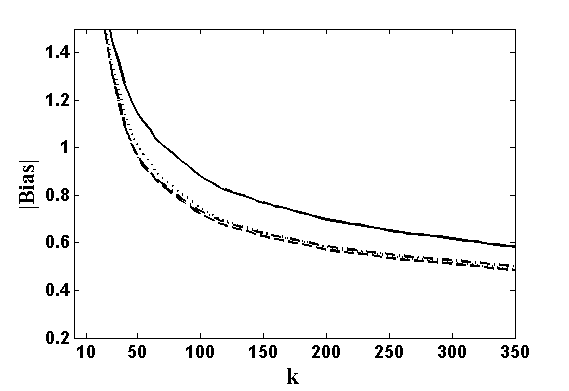}
\includegraphics[width= 0.4\textwidth,height= 0.2\textwidth] {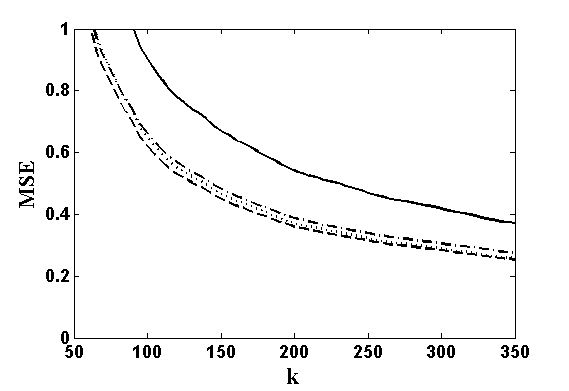}
\label{FIG:Sim_Cont_05_LN2}}
\\
\subfigure[$15\%$ contamination]{
\includegraphics[width= 0.4\textwidth,height= 0.2\textwidth] {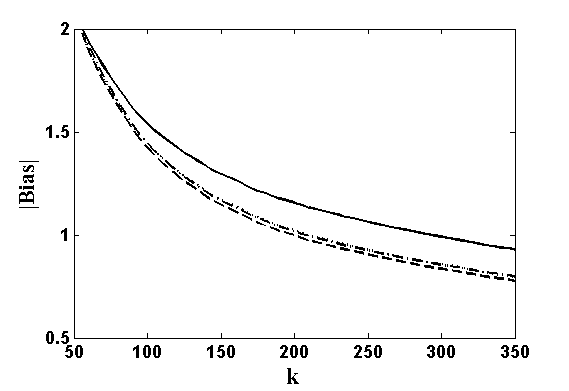}
\includegraphics[width= 0.4\textwidth,height= 0.2\textwidth] {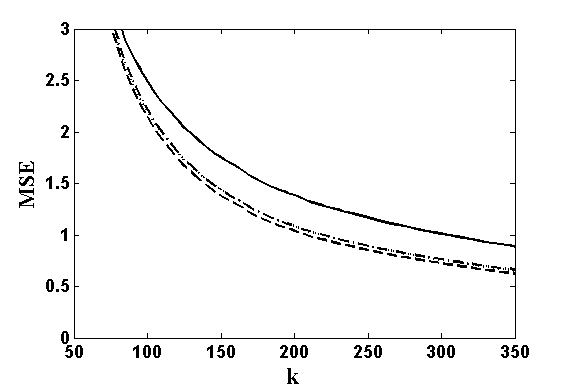}
\label{FIG:Sim_Cont_15_LN2}}
\caption{The empirical bias and MSE of the estimators $\hat{\gamma}_{ER,k}^{(\alpha)}$ 
 for model (M4) having $\gamma=0$ with contamination by the model (M1) having $\nu=1/3$ ($\gamma=3$) 
[Solid line: $\alpha=0$, Dotted line: $\alpha=0.3$, Dashed-dotted  line: $\alpha=0.5$, Dashed line: $\alpha=1$].}
\end{figure}

\clearpage\newpage
\begin{figure}[f]
\centering
\subfigure[$5\%$ contamination]{
\includegraphics[width= 0.4\textwidth,height= 0.2\textwidth] {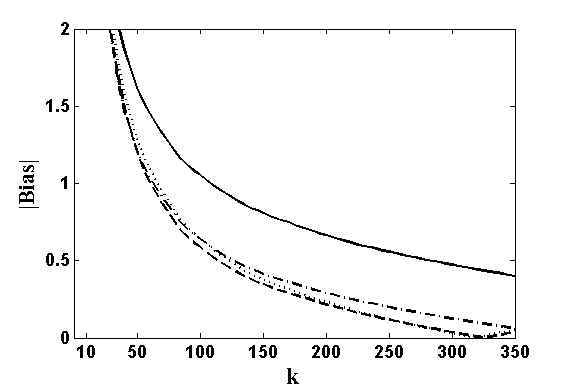}
\includegraphics[width= 0.4\textwidth,height= 0.2\textwidth] {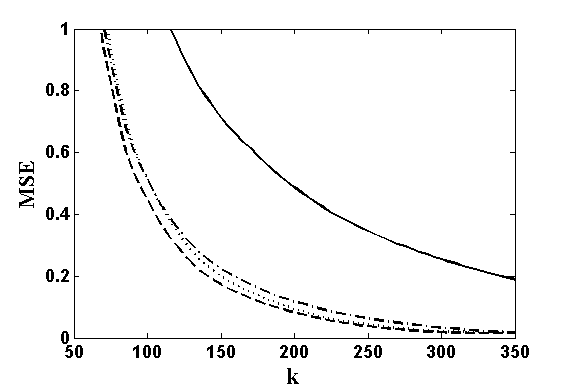}
\label{FIG:Sim_Cont_05_W2}}
\\
\subfigure[$15\%$ contamination]{
\includegraphics[width= 0.4\textwidth,height= 0.2\textwidth] {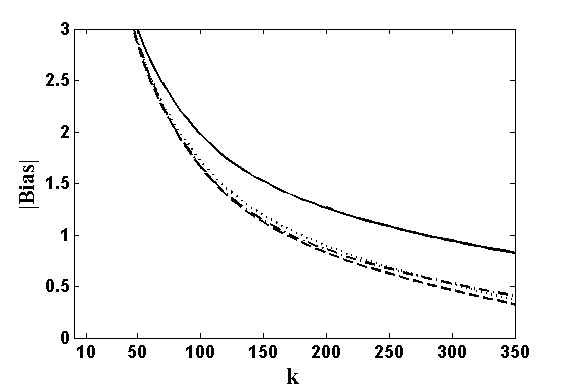}
\includegraphics[width= 0.4\textwidth,height= 0.2\textwidth] {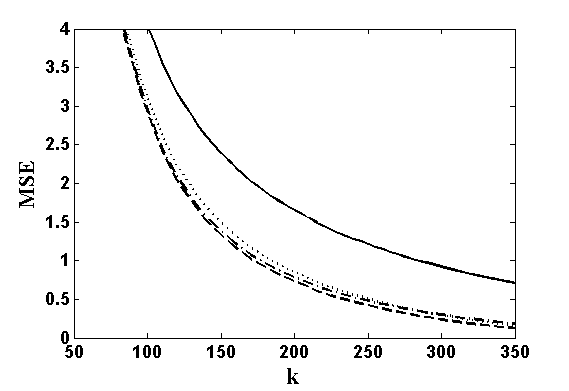}
\label{FIG:Sim_Cont_15_W2}}
\caption{The empirical bias and MSE of the estimators $\hat{\gamma}_{ER,k}^{(\alpha)}$ 
 for model (M5) having $\gamma=0$ with contamination by the model (M1) having $\nu=1/3$ ($\gamma=3$) 
[Solid line: $\alpha=0$, Dotted line: $\alpha=0.3$, Dashed-dotted  line: $\alpha=0.5$, Dashed line: $\alpha=1$].}
\end{figure}

\begin{figure}[f]
\centering
\subfigure[$5\%$ contamination]{
\includegraphics[width= 0.4\textwidth,height= 0.2\textwidth] {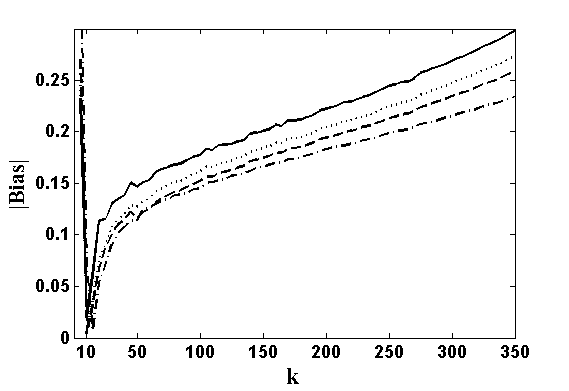}
\includegraphics[width= 0.4\textwidth,height= 0.2\textwidth] {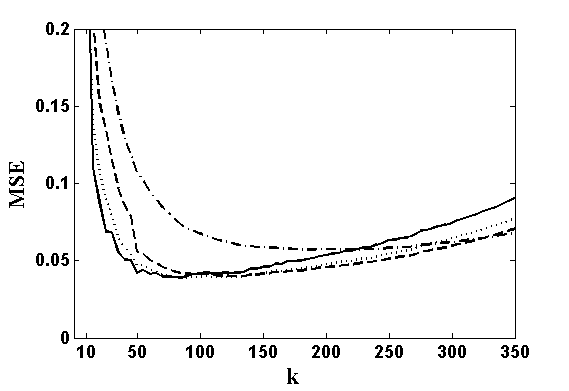}
\label{FIG:Sim_Cont_05_W3}}
\\
\subfigure[$15\%$ contamination]{
\includegraphics[width= 0.4\textwidth,height= 0.2\textwidth] {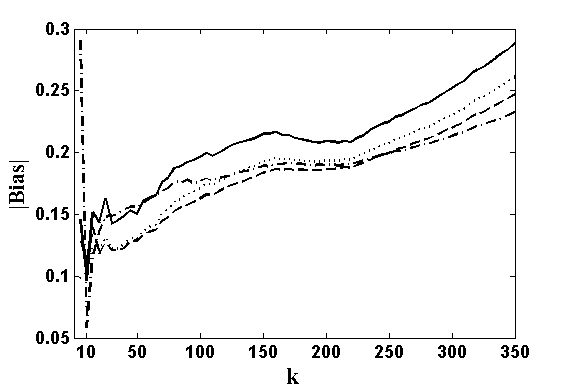}
\includegraphics[width= 0.4\textwidth,height= 0.2\textwidth] {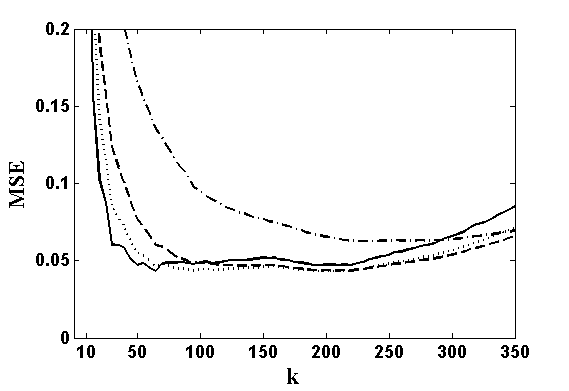}
\label{FIG:Sim_Cont_15_W3}}
\caption{The empirical bias and MSE of the estimators $\hat{\gamma}_{ER,k}^{(\alpha)}$ 
 for model (M5) having $\gamma=0$ with contamination by the model (M6) having $\gamma=-1$ 
[Solid line: $\alpha=0$, Dotted line: $\alpha=0.3$, Dashed-dotted  line: $\alpha=0.5$, Dashed line: $\alpha=1$].}
\end{figure}

\begin{figure}[f]
\centering
\subfigure[$5\%$ contamination]{
\includegraphics[width= 0.4\textwidth,height= 0.2\textwidth] {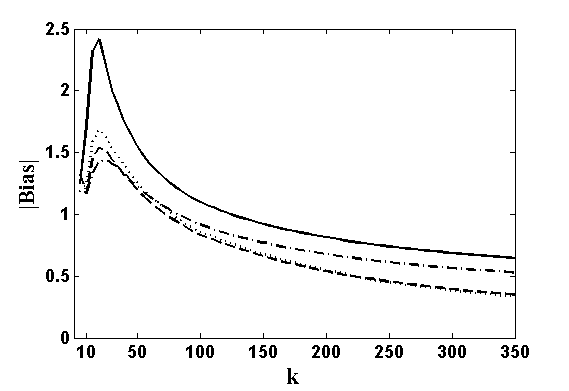}
\includegraphics[width= 0.4\textwidth,height= 0.2\textwidth] {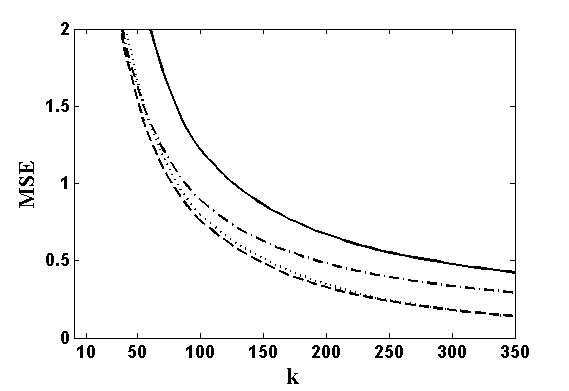}
\label{FIG:Sim_Cont_05_U1}}
\\
\subfigure[$15\%$ contamination]{
\includegraphics[width= 0.4\textwidth,height= 0.2\textwidth] {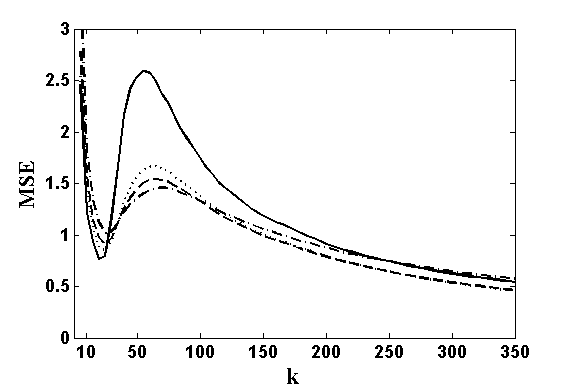}
\includegraphics[width= 0.4\textwidth,height= 0.2\textwidth] {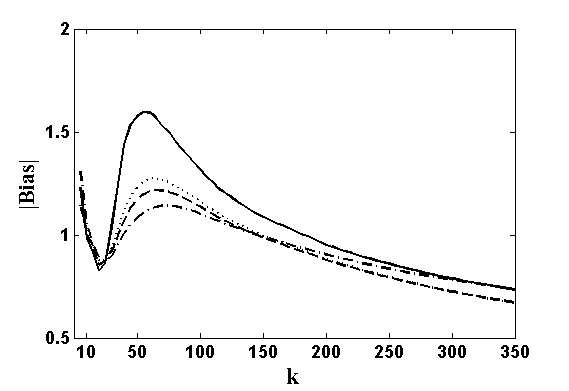}
\label{FIG:Sim_Cont_15_U1}}
\caption{The empirical bias and MSE of the estimators $\hat{\gamma}_{ER,k}^{(\alpha)}$
 for model (M6) having $\gamma=-1$ with contamination by the  model (M5) having $\gamma=0$
[Solid line: $\alpha=0$, Dotted line: $\alpha=0.3$, Dashed-dotted  line: $\alpha=0.5$, Dashed line: $\alpha=1$].}
\end{figure}

\begin{figure}[f]
\centering
\subfigure[$5\%$ contamination]{
\includegraphics[width= 0.4\textwidth,height= 0.2\textwidth] {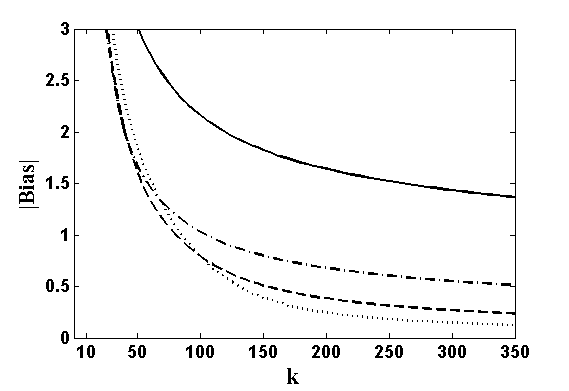}
\includegraphics[width= 0.4\textwidth,height= 0.2\textwidth] {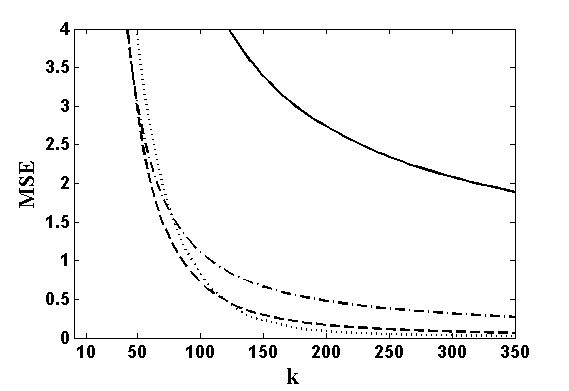}
\label{FIG:Sim_Cont_05_U2}}
\\
\subfigure[$15\%$ contamination]{
\includegraphics[width= 0.4\textwidth,height= 0.2\textwidth] {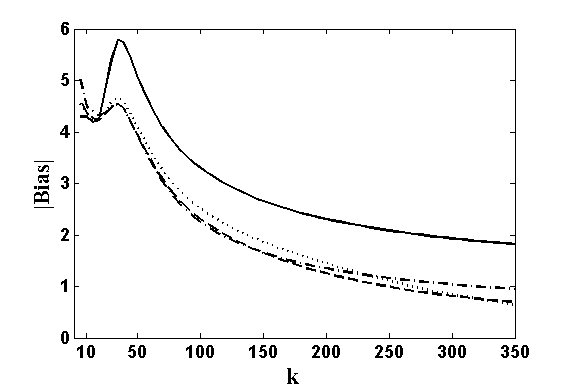}
\includegraphics[width= 0.4\textwidth,height= 0.2\textwidth] {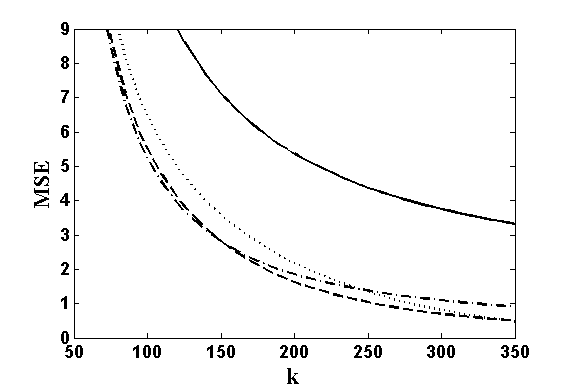}
\label{FIG:Sim_Cont_15_U2}}
\caption{The empirical bias and MSE of the estimators $\hat{\gamma}_{ER,k}^{(\alpha)}$
 for model (M6) having $\gamma=-1$ with contamination by the model (M1) having $\nu=1/3$ ($\gamma=3$) 
[Solid line: $\alpha=0$, Dotted line: $\alpha=0.3$, Dashed-dotted  line: $\alpha=0.5$, Dashed line: $\alpha=1$].}
\end{figure}

\begin{figure}[f]
\centering
\subfigure[$5\%$ contamination]{
\includegraphics[width= 0.4\textwidth,height= 0.2\textwidth] {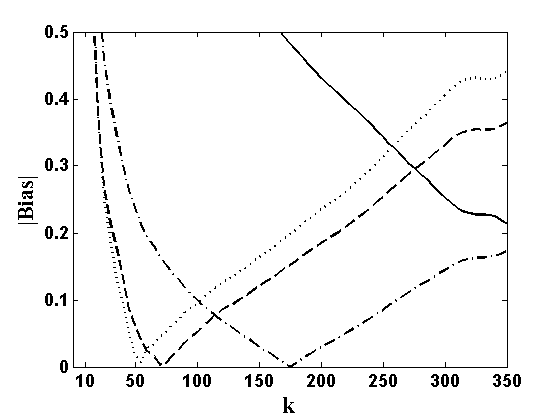}
\includegraphics[width= 0.4\textwidth,height= 0.2\textwidth] {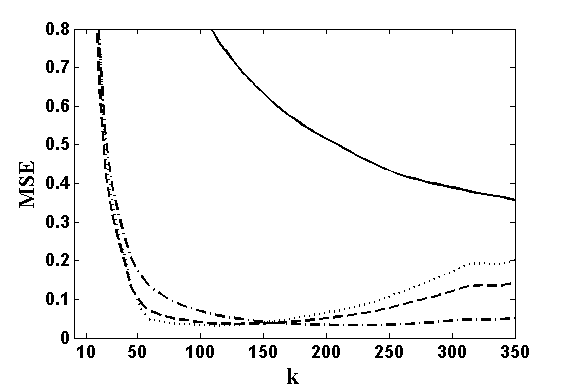}
\label{FIG:Sim_Cont_5_RBurr2}}
\\
\subfigure[$15\%$ contamination]{
\includegraphics[width= 0.4\textwidth,height= 0.2\textwidth] {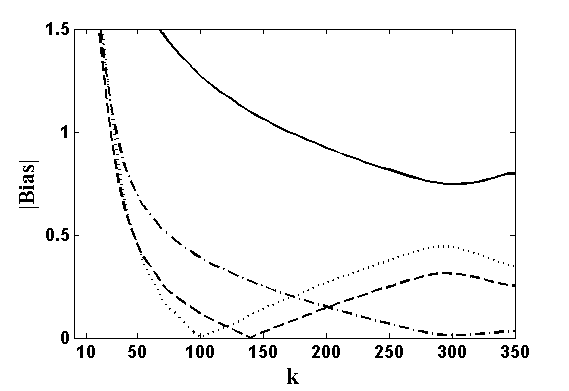}
\includegraphics[width= 0.4\textwidth,height= 0.2\textwidth] {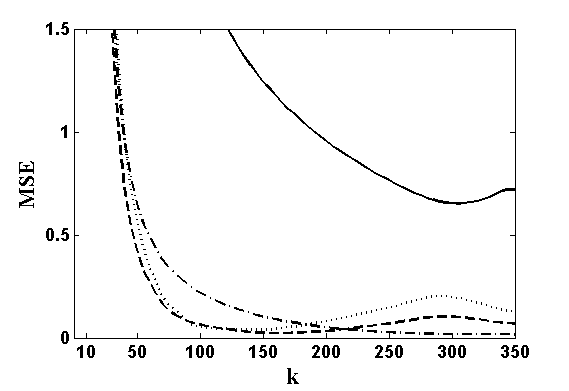}
\label{FIG:Sim_Cont_15_RBurr2}}
\caption{The empirical bias and MSE of the estimators $\hat{\gamma}_{ER,k}^{(\alpha)}$ 
 for model (M7) having $(\beta,~\tau,~\lambda) = (1,~1,~1)$  
with contamination by the model (M2) having $(\beta,~\tau,~\lambda) = (4,~0.25,~1)$  
[Solid line: $\alpha=0$, Dotted line: $\alpha=0.3$, Dashed-dotted  line: $\alpha=0.5$, Dashed line: $\alpha=1$].}
\end{figure}


\begin{thebibliography}{}


\bibitem{a12}
Alfons, A., Holzer, J. and Templ, M. (2013a) 
{\bf laeken}: {\it Estimation of Indicators on Social Exclusion and Poverty}. 
http://CRAN.R-project.org/package=laeken, R Package Version 0.4.4.


\bibitem{ak12}
Alfons, A., Kraft, S. (2012). 
{\bf simPopulation}: {\it Simulation of Synthetic Populations for Surveys based on Sample Data.}
http://CRAN.R-project.org/package=simPopulation, R Package version 0.4.0.

\bibitem{aktf11}
Alfons, A., Kraft, S., Templ, M. and Filzmoser, P. (2011a).
Simulation of close-to-reality population data for household surveys with application to EU-SILC. 
{\it Statist. Meth. Applic.}, {\bf 20(3)}, 383--407.


\bibitem{at13}
Alfons, A., and Templ, S. (2013).
Estimation of social exclusion indicators from complex surveys: The R package leaken. 
{\it Journal of Statistical Software}, {\bf 54(15)}, 1--25


\bibitem{aktf11b}
Alfons, A., Templ, M., Filzmoser, P., and Holzer, J. (2011b).
Robust Pareto tail modeling for the Estimation of Indicators on Social Exclusion using the R Package laeken. 
Research Report CS-2011-2, Department of Statistics and Probability Theory, Vienna University of Technology.


\bibitem{at13}
Alfons, A., Templ, M. and Filzmoser, P. (2013b).
Robust estimation of economic indicators from survey samples based on Pareto tail modelling.
{\it Journal of Royal Statistical Society, Series C}, {\bf 62(2)}, 271--286.


\bibitem{bhhj98}
Basu, A., Harris, I. R., Hjort, N. L. and Jones, M. C. (1998). 
Robust and efficient estimation by minimising a density power divergence. 
{\it Biometrika}, {\bf 85}, 549--559.

\bibitem{bdgm99}
Beirlant, J., Dierckx, G., Goegebeur, Y. and Matthys, G. (1999). 
Tail index estimation and an exponential regression model.
{\it Extremes}, {\bf 2(2)}, 177--200.

\bibitem{bdgm96}
Beirlant, J., Vynckier, P. and Teugels, J. L. (1996). 
Tail index estimation, Pareto quantile plots, and regression diagnostics.
{\it Journal of the American Statistical Association}, {\bf 31(436)}, 1659--1667.


\bibitem{deh89}
Dekkers, A.~L.~M., Einmahl, J.~H.~J. and de Haan, L. (1989). 
A moment estimator for the index of an extreme value distribution. 
{\it Ann. Statist.}, {\bf 17}, 1833--1855.


\bibitem{gb13}
Ghosh, A. and Basu, A.~(2013). 
Robust Estimation for Independent Non-Homogeneous Observations using Density Power Divergence with Applications 
to Linear Regression. 
{\it Electronic Journal of statistics}, {\bf 7}, 2420--2456.

\bibitem{g43}
Gnedenko, B.~V. (1943). 
Sur la distribution limite du terme maximum d'une s\'{e}rie al\'{e}atoire.
{\it Ann. Math.}, {\bf 44}, 423--453.

\bibitem{ggr14}
Goegebeur, Y.,  Guillou, A. and Rietsch T. (2014). 
Robust conditional Weibull-type estimation.
{\it Ann. Inst. of Statist. Math.}, 1--36.


\bibitem{dh70}
de Haan, L. (1970). 
On regular variation and its application to the weak convergence of sample extremes. 
{\it Math. centre Tract}, {\bf 32}, Amsterdam.

\bibitem{Hampel:1968}
Hampel, F.~R. (1968).
\newblock {\em Contributions to the theory of robust estimation}.
\newblock Ph.\ D. thesis, University of California, Berkeley, USA.

\bibitem{Hampel:1974}
Hampel, F.~R. (1974).
\newblock The influence curve and its role in robust estimation.
\newblock {\em J. Amer. Statist Assoc.\/}~{\em 69}, 383--393.


\bibitem{h75}
Hill, B.~M. (1975). 
A simple general approach to inference about the tail of a distribution.
{\it Ann. Statist.}, {\bf 3}, 1163--1174.


\bibitem{kl08}
Hosking, J.~R.~M. and Wallis, J.~R. (1987). 
Parameter and quantile estimation for the generalized Pareto distribution.
{\it Technometrics}, {\bf 29}, 339--349.

\bibitem{h11}
Hulliger, B., Alfons, A., Bruch, C., Filzmoser, P., Graf, M., Kolb, J.-P., Lehtonen, R., Lussmann, D., Meraner,
A., Münnich, R., Nedyalkova, D., Schoch, T., Templ, M., Valaste, M., Veijanen, A. and Zins, S. (2011).
Report on the simulation results: Deliverable D7.1, AMELI Project.


\bibitem{kl08}
Kim, M. and Lee, S. (2008). 
Estimation of a tail index based on minimum density power divergence.
{\it Journal of multivariate Analysis}, {\bf 99}, 2453--2471.

\bibitem{my02}
Marazzi, A. and Yohai, V. (2004). 
Adaptively truncated maximum likelihood regression with asymmetric errors.
{\it Jounal of Statistical Planning and Inference}, {\bf 122}, 271--291.


\bibitem{mb03}
Matthys, G. and Beirlant, J. (2003). 
Estimating the extreme value index and high quantiles with exponential regression models.
{\it Statistica Sinica}, {\bf 13}, 853--880.


\bibitem{p13}
Pak R. J. (2013). 
A robust estimation for the composite lognormal-Pareto model.
{\it Communications for statistical Applications and methods}, {\bf 20(4)}, 311--320.


\bibitem{p75}
Pickands III, J. (1975). 
Statistical inference using extreme order statistics.
{\it Ann. Statist.}, {\bf 3}, 119--131.


\bibitem{s87}
Smith, R.~L. (1987). 
Estimating tails of probability distributions.
{\it Ann. Statist.}, {\bf 15}, 1174--1207.


\bibitem{Stigler:1977}
Stigler, S.~M. (1977).
\newblock Do robust estimators work with real data?
\newblock {\em Annals of Statistics}, {\bf 5}, 1055--1098.


\bibitem{vbh04}
Vandewalle, B., Beirlant, J. and Hubert, M. (2004). 
A robust estimator of the tail index based on an exponential regression model. 
In {\it Theory and Applications of Recent Robust Methods}, 
Editors M. Hubert, G. Pison, A. Struyf, and S. van Aelst.
367-376. Basel:Birkhauser.


\bibitem{vbch07}
Vandewalle, B. Beirlant, J. Christmann, A. and Hubert, M. (2007). 
A robust estimator for the tail index of Pareto-type distributions. 
{\it Computational Statistics \& Data Analysis}, {\bf 51(12)}, 6252--6268.



\end{thebibliography}
\end{document}